\documentclass[aps,prb,reprint,groupedaddress]{revtex4-2}

\usepackage{amsmath}
\usepackage{amsthm}
\usepackage{amssymb}
\usepackage{graphicx}
\usepackage{tikz}
\usepackage{tikz-feynman}
\usepackage{breqn}
\usepackage{hyperref}
\usepackage{xcolor}
\usepackage[normalem]{ulem}

\newcommand{\ket}[1]{| #1 \rangle}

\newcommand{\avg}[1]{\langle #1 \rangle}

\newcommand{\op}[1]{\hat{ #1 }}

\makeatletter
\let\cat@comma@active\@empty
\makeatother

\begin{document}

\title{Dynamical mean-field approach to disordered interacting systems and applications to quantum transport problem}
\author{Jiawei Yan}
\email{jiawei.yan@unifr.ch}
\affiliation{Department of Physics, University of Fribourg, 1700 Fribourg, Switzerland}
\author{Philipp Werner}
\email{philipp.werner@unifr.ch}
\affiliation{Department of Physics, University of Fribourg, 1700 Fribourg, Switzerland}

\date{\today}

\begin{abstract}
We discuss a non-equilibrium dynamical mean-field framework for simulating inhomogeneous Hubbard models with local disorders.
Our approach treats electron interactions and disorders on equal footing, by considering only local dynamical fluctuations.
The theory reduces to non-equilibrium dynamical mean-field theory in the presence of only electron-electron interactions and to the coherent potential approximation in noninteracting systems with disorders.
Both time-dependent and steady-state problems are treated by implementing the theory on the three branch Kadanoff-Baym contour and two-branch Keldysh contour, respectively.
Benchmarks on a $8$-site cube show that the method yields rather accurate spectral functions in both the weakly and strongly interacting regimes.
In a cubic lattice, we demonstrate energy conservation after an interaction quench and thermalization after just a few hopping times in both clean and disordered systems.
As an application, we study transport through a serial double quantum-dot sandwiched between two leads, focusing on the current and dot occupations after a voltage quench.
\end{abstract}

\maketitle

\section{Introduction\label{sec: introduction}}

The interplay between electron-electron (el-el) interactions and disorders plays an important role in many widely studied condensed matter phenomena, including metal-insulator transitions  \cite{Belitz1994,Lee1985,N.F.1968}, superconductivity \cite{Lee2006}, giant magnetoresistance \cite{Parkin2004}, and many-body localization \cite{Abanin2019}. In the context of quantum transport, this physics also influences the behavior of devices, which is exploited, e.~g., through functionalized chemical doping \cite{Zwanenburg2013}.
Consequently, there is a need to develop computational methods that qualitatively or even quantitatively capture the combined effect of el-el interactions and disorders, both in equilibrium and non-equilibrium setups.

The formulation of a microscopic theory that involves both el-el interactions and disorder degrees of freedom is challenging. 
The challenges originate primarily from two factors:
(i) strong Coulomb interactions correlate the motion of the electrons and prevent the use of effective single-particle descriptions \cite{Bruus2004,Mahan2000};
(ii) the presence of disorders breaks the translational invariance of the system, so that Bloch theory becomes invalid \cite{Okhotnikov2016}.
Both effects lead to an exponential scaling of the complexity of the problem with system size, so that exact results can be obtained only for very small systems.
Moreover, when considering quantum transport problems, the proper nonequilibrium distribution of the occupied states needs to be taken into account \cite{Datta2005,*Datta1997}.

Over the past decades, various computational methods have been developed to address the challenges posed by the correlated electron \cite{Gubernatis2016,Schollwoeck2005,Caffarel1994,Bruus2004} and disorder problem \cite{gonis1992green,Zunger1990}.
Among them, dynamical mean-field theory (DMFT) stands out due to its non-perturbative nature, its possible combination with density functional theory input for the simulation of real materials, and its natural extension to non-equilibrium conditions \cite{Georges1996, Kotliar2006, PhysRevLett.97.266408,RevModPhys.86.779}.
The fundamental idea of DMFT is to map the original lattice onto an auxiliary impurity problem (small correlated system coupled to a non-interacting bath).
The bath of this auxiliary problem is self-consistently determined and mimics the effect of the lattice environment. 
DMFT is exact in both the infinite dimensional and atomic limits, and provides a reasonable interpolation between them \cite{Metzner1989}.
In studies of noninteracting systems with disorders, the method is also known as coherent potential approximation (CPA).
The CPA was initially formulated by introducing a coherent medium, which is self-consistently determined by the condition that the averaged on-site scattering of any given site vanishes when embedded in the effective medium \cite{Soven1967,Velicky1968,Elliott1974}. 
A deeper understanding of CPA emerged with the development of a functional integral formulation, which revealed that CPA is a special case of DMFT for disordered systems \cite{Jani1989}.
This connection provides a solid basis for combining the two methods to address systems that involve both el-el interactions and disorders.

This idea was initially introduced in Ref.~\cite{PhysRevB.46.15712,*Janis1993}, where the disorder averaged free energy functional of the Hubbard-Anderson model was derived in the infinite-coordination limit.
Subsequently, magnetic phase diagrams and Mott-Anderson transitions were investigated on this (dynamical) mean-field level \cite{Ulmke1995,PhysRevLett.78.3943}.
Combinations with first-principles methods, which enable the simulation of equilibrium properties of real materials,  were also reported \cite{Drchal1999,*Ebert2011}. 
Further efforts have been made to account for non-local spatial fluctuations by combining the theory with the dual fermion approach \cite{Terletska2013,*Yang2014} and by incorporating off-diagonal disorder using the Blackman-Esterling-Berk transformation \cite{Weh2021}.

Very recently, a nonequilibrium extension of the combined CPA and DMFT approach was presented in Ref.~\cite{PhysRevB.106.195156} and applied to an interaction quench problem on the Bethe lattice.
In this work, we use the same method to study disordered interacting systems on an inhomogeneous Hubbard-Anderson lattice.
We formulate the theory both in terms of nonequilibrium Green's functions defined on a three-branch Kadanoff-Baym (KB) contour, appropriate for simulations starting from an equilibrium state, and with real-frequency Green's functions for the simulation of nonequilibrium steady-states. 
In contrast to the previous study (Ref.~\cite{PhysRevB.106.195156}), which considered a Bethe lattice, we formulate the method for generic finite-dimensional lattices.
This enables us to benchmark the method against exact diagonalization results on small isolated systems.
Additionally, in our implementation, Langreth's rules \cite{stefanucci2013nonequilibrium,SCHULER2020107484} are applied to ensure the causal structure of the time propagation and the implementation of high-order discretization schemes in the numerical implementation.

On the application side, we mainly focus on a quantum transport setup consisting of a central scattering region with both disorders and interactions which is coupled to two metallic leads. We account for disorders in the local energy and in the local el-el repulsion, and drive the system out of equilibrium by a voltage quench.  

The article is structured as follows. In Sec.~\ref{sec: theory}, we present the nonequilibrium DMFT based formalism.
Specifically, we discuss the self-consistency loop, the impurity solver used, the calculation of configurationally averaged physical observables and the implementation for both time-dependent and steady-state calculations. 
In Sec.~\ref{sec: results}, we discuss the numerical results, including the equilibrium spectral function of an $8$-site cube, an interaction quench problem on a cubic lattice, and a serial double quantum dots system coupled to two external leads under a step-shaped voltage profile.
We conclude in Sec.~\ref{sec: conclusions}, while detailed derivations of the theory are provided in the appendices.

\section{Theory\label{sec: theory}}

We study a single-orbital Hubbard model, given by the Hamiltonian
\begin{dmath}\label{eq: model hamiltonian}
\op{H}(t) = \sum_{i,\sigma} (\epsilon_{i\sigma}(t) - \mu) c_{i\sigma}^\dag c_{i\sigma} + \sum_{i\ne j,\sigma} W_{ij,\sigma}(t) c_{i\sigma}^\dag c_{j\sigma} +\sum_i U_i(t) \hat{n}_{i\uparrow} \hat{n}_{i\downarrow}~,
\end{dmath}
where $c_{i\sigma}^\dag$ and $c_{i\sigma}$ are the creation and annihilation operators for an electron located at site $i$ with spin $\sigma$, $\hat{n}_{i\sigma} = c_{i\sigma}^\dag c_{i\sigma}$ is the electron number operator, $\epsilon_{i\sigma}(t)$ and $W_{ij,\sigma}(t)$ are the local energies and hopping integrals, $U_{i}(t)$ is the on-site Coulomb integral, and $\mu$ is the chemical potential.
Due to the hermiticity of the Hamiltonian, the hoppings satisfy $W_{ij,\sigma}(t) = W_{ji,\sigma}^*(t)$. 

We furthermore consider an ensemble of disorder configurations, where each lattice site can be in a configuration $Q \in \{A,B,\dots\}$. 
The probability of site $i$ to be in configuration $Q$ is denoted by $p_i^Q$, and the  configurations on different sites are assumed to be uncorrelated. 
Only $\epsilon_{i\sigma}(t)$ and $U_i(t)$ depend on the disorder configuration, i.e., they take the values $\epsilon_{i\sigma}^Q(t)$ and $U_i^Q(t)$, while 
$W_{ij,\sigma}(t)$ is not affected by the disorder.

\subsection{Dynamical mean-field theory formulation}

\begin{figure}
\begin{tikzpicture}[scale=.4]
	\node (0) at (0, 1) {};
	\node (1) at (10, 1) {};
	\node (2) at (0, -1) {};
	\node (3) at (10, -1) {};
	\node (4) at (0, -3.8) {};
	\node (5) at (-1.5, 0) {};
	\node (6) at (14, 0) {};
	\node (7) at (0, -.5) {$t=t_0$};
	\node (8) at (12.5, -.5) {$t=t_f$};
	\node (9) at (0, -4.2) {$t=t_0-i\beta$};
	\draw [very thick] (0.center) to (1.center);
	\draw [very thick, bend left=90, looseness=1.75] (1.center) to (3.center);
	\draw [very thick] (3.center) to (2.center);
	\draw [very thick, ->] (2.center) to (4.center);
	\draw [dashed, very thick, ->] (5.center) to (6.center);
\end{tikzpicture}
\caption{Schematic illustration of the Kadanoff-Baym contour $\mathcal{C}$ in the complex time plane. Here, $\beta$ is the inverse temperature of the initial state. 
\label{fig: Kadanoff-Baym contour}}
\end{figure}
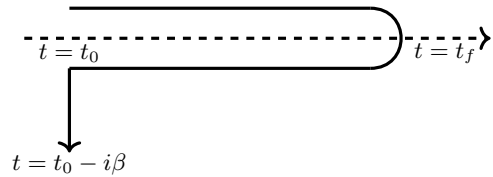

Our theory is formulated on the three-branch Kadanoff-Baym contour $\mathcal{C}$, which is used to describe systems that are initially (at time $t_0$) prepared in equilibrium at inverse temperature $\beta$ and subsequently driven out of equilibrium by external fields \cite{stefanucci2013nonequilibrium}. 
The contour starts at $t_0$, extends up to a maximum time of interest $t_f$ along the real-time axis, then returns to $t_0$, and finally extends along the imaginary-time axis to $-i\beta$, see illustration in Fig. \ref{fig: Kadanoff-Baym contour}. 
The single-particle Green's function for a specified disorder configuration is given by \cite{stefanucci2013nonequilibrium,Haug2008,Kamenev2011}
\begin{dmath}\label{eq: contour-ordered Green's function}
G_{ij,\sigma}(z,z') = -i \avg{c_{i\sigma}(z) c_{j\sigma}^*(z')}_{S^{\text{lat}}}~,
\end{dmath}
where $z$ denotes a time argument on the contour and $c$ ($c^*$) denote Grassmann variables for the $c$-electrons. 
(We use the same notation as for the creation and annihilation operator, since they can be distinguished from the context.)
$\avg{\cdots}_{S} = \frac{1}{Z} \int_{\mathcal{C}} \cdots \mathcal{D}[c^*,c] e^{iS}$ is the expectation value for a given action $S$, with $Z = \int_{\mathcal{C}} \mathcal{D}[c^*,c] e^{iS}$ the partition function of the initial state.
 $S^{\text{lat}}$ in Eq.~\eqref{eq: contour-ordered Green's function} is the lattice action for a specific disorder configuration,
 \begin{dmath}\label{eq: lattice action}
S^{\text{lat}} = \int_{\mathcal{C}} dz
\left\{ \sum_{ij\sigma} c_{i\sigma}^*(z) \left[ \delta_{ij} \left( i\frac{\overrightarrow{d}}{dz} + \mu - \epsilon_{i\sigma}(z)\right) - W_{ij,\sigma}(z) \right]  c_{j\sigma}(z)
- \sum_i U_i(z) n_{i\uparrow}(z) n_{i\downarrow}(z) \right\}
- \int_{\mathcal{C}} dz dz' \sum_{ij\sigma} c^*_{i\sigma}(z) \Sigma^{\text{ext}}_{ij,\sigma}(z,z') c_{j\sigma}(z')~.
\end{dmath}
Here, the time arguments of the model parameters $\epsilon$, $W$ and $U$ are extended to the complex plane \cite{stefanucci2013nonequilibrium}. 
We also introduced a generic non-hermitian bilinear source $\Sigma^{\text{ext}}_{ij,\sigma}(z,z')$ in Eq.~\eqref{eq: lattice action}. In a quantum transport set-up, this source term can be used to represent the effect of external leads that drive the system out of equilibrium.
The calculation of the lead self-energy $\Sigma^{\text{ext}}$ is discussed in Appendix~\ref{sec: lead self-energy}.

Since we are considering an ensemble of disorder configurations, the free energy (generating functional) reads $\Omega = -\frac{1}{\beta} \avg{ \ln Z}_{\text{dis}}$, where $\avg{\cdots}_{\text{dis}}$ refers to the ensemble average over the disorder configurations.
The disorder averaged lattice Green's function can formally be expressed as
\begin{dmath}
\avg{G_{ij,\sigma}(z,z')}_{\text{dis}} = \beta \frac{\delta \Omega[\Sigma^{\text{ext}}]}{\delta \Sigma_{ji,\sigma}^{\text{ext}}(z',z)}~.
\end{dmath}
In the infinite-dimensional limit $d \rightarrow \infty$, with the hopping parameters rescaled as $W_{ij}(t) \rightarrow W_{ij}(t) / d^{|i-j|/2}$ \cite{Metzner1989}, the solution of the lattice problem \eqref{eq: lattice action} reduces to the solution of impurity problems for the different $Q$, with action
\begin{dmath}\label{eq: local action}
S_j^{Q,\text{imp}} = \int_{\mathcal{C}} dz\left\{ \sum_{\sigma} c_{j\sigma}^*(z) \left( i\frac{\overrightarrow{d}}{d z} + \mu - \epsilon^Q_{j\sigma}(z) \right) c_{j\sigma}(z) - U^Q_{j}(z) c_{j\uparrow}^*(z) c_{j\uparrow}(z) c_{j\downarrow}^*(z) c_{j\downarrow}(z)\right\}
- \int_{\mathcal{C}} dz dz' \sum_{\sigma} c_{j\sigma}^*(z) \Delta^{\text{imp}}_{j\sigma}(z,z') c_{j\sigma}(z')~,\label{action_Q}
\end{dmath}
and the disorder average reduces to an average over a single site \cite{PhysRevB.46.15712,Janis1993}.
In Eq.~\eqref{eq: local action}, $\Delta^{\text{imp}}_{j\sigma}(z,z')$ is the impurity hybridization function of site $j$, describing the amplitude for hopping from site $j$ into the rest of the lattice at time $z'$ and returning back to site $j$ at time $z$.
Note that $\Delta^{\text{imp}}_{j\sigma}(z,z')$ is independent of the species $Q$ on site $j$.
From the impurity actions (\ref{action_Q}), the impurity Green's functions can be calculated as
\begin{dmath}\label{eq: impurity Green's function}
G_{j\sigma}^{Q,\text{imp}}(z,z') = -i\avg{ c_{j\sigma}(z) c_{j\sigma}^*(z') }_{S_j^{Q,\text{imp}}}~,
\end{dmath}
and the disorder average of these impurity Green's functions yields the averaged local lattice Green's function.

The same procedure can be applied to a finite-dimensional system, which corresponds to the dynamical mean field theory (DMFT) approximation \cite{Georges1996}.
To derive the formalism, we first introduce a (exact) non-interacting effective medium, whose properties are governed by the action
 \begin{dmath}\label{eq: effective medium action}
S^{\text{eff},\star} = \int_{\mathcal{C}} dz dz'
\left\{ \sum_{ij\sigma} c_{i\sigma}^*(z) \left[ \delta(z-z') \delta_{ij} \left( i\frac{\overrightarrow{d}}{dz} + \mu \right) - \tilde{W}_{ij,\sigma}(z,z') - \Sigma^{\text{eff},\star}_{ij,\sigma}(z,z') \right]  c_{j\sigma}(z') \right\}~,
\end{dmath}
where $\tilde{W}_{ij,\sigma}(z,z') = W_{ij}(z) \delta(z-z') +\Sigma_{ij,\sigma}^{\text{ext}}(z,z')$. 
In Eq.~\eqref{eq: effective medium action}, $\Sigma^{\text{eff},\star}$ is the self-energy of the effective medium, which in general is non-local in both space and time. 
Note that in contrast to $S^{\text{lat}}$ in Eq.\eqref{eq: lattice action} which depends on the disorder configuration, $S^{\text{eff},\star}$ is without randomness since the effects of the local Coulomb interaction and onsite energy have been absorbed into $\Sigma^{\text{eff},\star}$.
The Green's function of the effective medium, $\Gamma_{ij,\sigma}^\star(z,z') = -i\avg{c_{i\sigma}(z) c_{j\sigma}^*(z')}_{S^{\text{eff},\star}}$, is supposed to reproduce the disorder averaged Green's function of the interaction lattice, $\avg{G_{ij,\sigma}(z,z')}_{\text{dis}}$.
This identity provides a formal definition of the exact effective self-energy $\Sigma^{\text{eff},\star}$.

Evaluating $\Sigma^{\text{eff},\star}$ is very costly because of the exponential scaling of the many-body Hilbert space and the disorder configurational space with increasing number of lattice sites. 
To make such calculations feasible, one can employ the DMFT approximation.
The idea is to retain in Eq.~\eqref{eq: effective medium action} only the local time-dependent fluctuations from the interactions and disorders, and neglect all the spatial fluctuations.
This corresponds to the approximation $\Sigma_{ij,\sigma}^{\text{eff},\star}(z,z') \approx \delta_{ij} \Sigma_{jj,\sigma}^{\text{eff}}(z,z')$, which becomes exact in the infinite dimensional limit.
The action for the approximated effective medium reads
 \begin{dmath}\label{eq: approximated effective medium action}
S^{\text{eff}} = \int_{\mathcal{C}} dz dz' \sum_{ij\sigma} c_{i\sigma}^*(z) \left\{  \delta_{ij} \left[ \delta(z-z') \left( i\frac{\overrightarrow{d}}{dz} + \mu \right) - \Sigma_{jj,\sigma}^{\text{eff}}(z,z') \right] - \tilde{W}_{ij,\sigma}(z,z') \right\}  c_{j\sigma}(z') ~.
\end{dmath}
In the above equation, $\Sigma_{jj,\sigma}^{\text{eff}}(z,z')$ can be viewed as the non-hermitian atomic level of the effective lattice, also known as the coherent potential in the CPA community \cite{turek2013electronic, PhysRevB.94.045424}. 
This coherent potential should be calculated self-consistently.
The Green's function corresponding to $S^{\text{eff}}$ will be denoted by $\Gamma_{ij,\sigma}(z,z')$ in the following, 
\begin{dmath}\label{eq: lattice Green's function}
\Gamma_{ij,\sigma}(z,z') = -i\avg{c_{i\sigma}(z) c_{j\sigma}^*(z')}_{S^{\text{eff}}}~.
\end{dmath}

To formulate the self-consistent loop which determines the effective medium, it is useful to introduce the locator $\gamma_j$ as the Green's function of the effective medium in the atomic limit
\begin{subequations}\label{eq: coherent locator}
\begin{dmath}
\left(+i\frac{\overrightarrow{d}}{dz} + \mu \right) \gamma_{j\sigma}(z,z') = \delta(z-z') + [ \Sigma_{jj,\sigma}^{\text{eff}} * \gamma_{j\sigma}] (z,z')~,
\end{dmath}
\begin{dmath}
\gamma_{j\sigma}(z,z') \left(-i\frac{\overleftarrow{d}}{dz'} + \mu \right) = \delta(z-z') +  [\gamma_{j\sigma} * \Sigma_{jj,\sigma}^{\text{eff}} ] (z,z')~,
\end{dmath}
\end{subequations}
where $[A*B](z,z') = \int_{\mathcal{C}} d\bar{z} A(z,\bar{z}) B(\bar{z},z')$ denotes the convolution on the $\mathcal{C}$-contour.
From Eq.~\eqref{eq: coherent locator} it follows that $\Sigma^{\text{eff}}$ and $\gamma$ are in one-to-one correspondence, and thus either of the two functions can be used to characterize the effective medium. In the following we proceed with $\gamma$. 
From Eqs.~\eqref{eq: approximated effective medium action},\eqref{eq: lattice Green's function} and \eqref{eq: coherent locator} one obtains the lattice Dyson equation
\begin{dmath}\label{eq: lattice Dyson equation}
\Gamma_{ij,\sigma}(z,z') = \gamma_{i\sigma}(z,z')\delta_{ij} + [\gamma_{\sigma} * \tilde{W}_{\sigma} * \Gamma_{\sigma}]_{ij}(z,z')~.
\end{dmath}
By iterating Eq.~\eqref{eq: lattice Dyson equation}, the local components of $\Gamma$ can be expressed as
\begin{dmath}\label{eq: impurity dyson equation}
\Gamma_{jj,\sigma}(z,z') =  \gamma_{j\sigma}(z,z') + [\gamma_{j\sigma} * \Delta^{\text{lat}}_{j\sigma} * \Gamma_{jj,\sigma}](z,z')~,
\end{dmath}
where
\begin{dmath}\label{eq: lattice hybridization function}
\Delta^{\text{lat}}_{j\sigma}(z,z') = 
\tilde{W}_{jj,\sigma}(z,z')
+ \sum_{l\neq j} [\tilde{W}_{jl,\sigma} * \gamma_{l\sigma} * \tilde{W}_{lj,\sigma}](z,z')
+ \sum_{l,m\neq j} [\tilde{W}_{jl,\sigma} * \gamma_{l\sigma} * \tilde{W}_{lm,\sigma} * \gamma_{m\sigma} * \tilde{W}_{mj,\sigma}](z,z')
+ \sum_{l,m,n\neq j}
[\tilde{W}_{jl,\sigma} * \gamma_{l\sigma} * \tilde{W}_{lm,\sigma} * \gamma_{m\sigma} * \tilde{W}_{mn,\sigma} * \gamma_{n\sigma} * \tilde{W}_{nj,\sigma}](z,z')
+ \cdots
\end{dmath}
is the lattice hybridization function, which accounts for all the scattering events in the effective medium that start and end on site $j$ but exclude any intermediate scattering processes involving site $j$ \cite{gonis1992green}.
For this reason, $\Delta_{j\sigma}^{\text{lat}}(z,z')$ is independent of the local occupation on site $j$.
The derivation of Eq.~\eqref{eq: lattice hybridization function} is given in Appendix \ref{sec: lattice hybridization function}.
In practice, Eq.~\eqref{eq: lattice hybridization function} can be recast into the form~\cite{RevModPhys.86.779}
\begin{dmath}\label{eq: lattice hybridization function 2}
\Delta_{j\sigma}^{\text{lat}}(z,z') + \int_{\mathcal{C}} d\bar{z} [\tilde{W}_\sigma * \Gamma_\sigma]_{jj}(z,\bar{z}) \Delta^{\text{lat}}_{j\sigma}(\bar{z},z') = \tilde{W}_{jj,\sigma}(z,z') + [\tilde{W}_\sigma * \Gamma_\sigma * \tilde{W}_\sigma]_{jj}(z,z')~,
\end{dmath}
which is a Volterra integro-differential equation of the second kind that we use in the numerical implementation \cite{SCHULER2020107484}.

To self-consistently determine the effective medium, i.e. $\gamma_{j\sigma}(z,z')$, one defines an impurity model (\ref{action_Q}) with hybridization function given by $\Delta^{\text{lat}}_{j\sigma}$:
\begin{dmath}\label{eq: scf condition 2}
\Delta_{j\sigma}^{\text{imp}}(z,z') = \Delta_{j\sigma}^{\text{lat}}(z,z')~.
\end{dmath}
In the following, we will use $\Delta$ to refer both to $\Delta^{\text{imp}}$ and $\Delta^{\text{lat}}$ without causing any ambiguities.
The self-consistency condition demands that the local lattice Green's function is equal to the disorder averaged impurity Green's function,
\begin{dmath}\label{eq: scf condition 1}
\avg{G_{j\sigma}^{Q,\text{imp}}(z,z')}_{\text{dis}} = \Gamma_{jj,\sigma}(z,z')~.
\end{dmath}
This identity allows to close the self-consistency loop.

The outlined formalism reduces to the conventional DMFT in the absence of disorder, while it reduces to CPA in the case of non-interacting systems.
On an infinitely connected Bethe lattice, the formalism becomes equivalent to the one presented in Ref.~\cite{PhysRevB.106.195156}.

In practice, the self-consistent calculation is performed by implementing the following steps:
(i)
One starts with an initial guess for the hybridization function $\Delta_{j\sigma}(z,z')$, for example $\Delta_{j\sigma}(z,z')=0$. 
(ii)
For given $\Delta_{j\sigma}(z,z')$ and configuration $Q$, one solves the impurity problem to obtain $G_{j\sigma}^{Q,\text{imp}}(z,z')$ defined in Eq.~\eqref{eq: impurity Green's function}, and then determines $\Gamma_{jj}(z,z')$  using Eq.~\eqref{eq: scf condition 1}.
The choice of the impurity solver is in principle arbitrary. The solver used in this work is described in the following subsection.
(iii)
With $\Delta_j(z,z')$ and $\Gamma_{jj}(z,z')$ fixed, Eq.~\eqref{eq: impurity dyson equation} is solved to obtain the locator $\gamma_j(z,z')$.
(iv)
The lattice Dyson equation \eqref{eq: lattice Dyson equation} is solved with the given $\gamma_j(z,z')$ to obtain the lattice Green's function $\Gamma_{ij,\sigma}(z,z')$.
(v)
The hybridization function is updated via Eq.~\eqref{eq: lattice hybridization function 2}.
Then steps (ii) to (v) are repeated until the hybridization function converges.

\subsection{Impurity solver: iterative perturbation theory}

A nontrivial problem in the self-consistency loop is the calculation of the impurity Green's function \eqref{eq: impurity Green's function} for the action \eqref{eq: local action}.
Calculating $G_{j\sigma}^{Q,\text{imp}}(z,z')$ is equivalent to evaluating the impurity self-energy $\Sigma_{j\sigma}^{Q,\text{imp}}(z,z')$, since these two functions are connected via the impurity Dyson equation (the conjugate equation is omitted)
\begin{dmath}
\left( +i\frac{\overrightarrow{d}}{dz} + \mu - \epsilon_{j\sigma}^{Q,\text{hf}}(z) - \Sigma^{Q,\text{imp}}_{j\sigma}(z,z') - \Delta_{j\sigma}(z,z') \right) G_{j\sigma}^{Q,\text{imp}}(z,z') = \delta(z-z')~.
\end{dmath}
where $\epsilon_{j\sigma}^{Q,\text{hf}}(z) = \epsilon_{j\sigma}^{Q}(z) + U^Q_j(z) n^Q_{j-\sigma}(z)$ and $\Sigma_{j\sigma}^{Q,\text{imp}}(z,z')$ is the dynamical self-energy that excludes the Hartree contribution.
Whether or not it is more convenient to calculate $G_{j\sigma}^{Q,\text{imp}}(z,z')$ or $\Sigma_{j\sigma}^{Q,\text{imp}}(z,z')$ depends on the impurity solver.
Here, we employ the iterated perturbation theory (IPT) \cite{Zhang1993}, which is computationally light, easy to extend to non-equilibrium situations, and qualitatively correct in the half-filled paramagnetic regime \cite{Tsuji2013}.

The impurity self-energy in IPT is given by
\begin{dmath}
\Sigma^{Q,\text{imp}}_{j\sigma}(z,z') = iU_j^Q(z) \mathcal{G}^{Q,\text{imp}}_{j\sigma}(z,z') \chi^{Q,\text{imp}}_{j-\sigma}(z,z') U_j^Q(z')~,
\end{dmath}
where $\mathcal{G}^{Q,\text{imp}}_{j\sigma}(z,z')$ is the (impurity) Weiss Green's function, satisfying
\begin{dmath}
\left( +i\frac{\overrightarrow{d}}{dz} + \mu - \epsilon_{j\sigma}^{Q,\text{hf}}(z) - \Delta_{j\sigma}(z,z') \right) \mathcal{G}_{j\sigma}^{Q,\text{imp}}(z,z') = \delta(z-z')~,
\end{dmath}
 and $\chi^{Q,\text{imp}}_{j-\sigma}(z,z') = -i \mathcal{G}^{Q,\text{imp}}_{j-\sigma}(z,z') \mathcal{G}^{Q,\text{imp}}_{j-\sigma}(z',z)$ is the electron-hole bubble.
Note that we build the IPT self-energy $\Sigma_{j\sigma}^{Q,\text{imp}}(z,z')$ from the Weiss Green's function, instead of the interacting impurity Green's function.
This implies that the IPT solution is not conserving in the Baym-Kadanoff sense \cite{PhysRev.127.1391,*PhysRev.124.287}.
However, previous studies showed that this variant gives quantitatively better results, compared to the self-consistent (boldified) IPT solution in short-time simulations \cite{Eckstein2010}.

\subsection{Physical observables}

Configurationally averaged physical observables can be obtained after the self-consistency loop has converged.
By construction (see Eq.~\eqref{eq: scf condition 1}), it does not matter if local one-particle quantities are calculated using the averaged impurity Green's function $G^{Q,\text{imp}}_{j\sigma}$ or the local lattice Green's function $\Gamma_{jj}$.

The electron density at site $j$ with spin $\sigma$ can be obtained from the lesser Green's function as
\begin{dmath}
n_{j\sigma}(t) = -i \Gamma_{jj,\sigma}^<(t,t)~.
\end{dmath}
The (probability) current flowing through the $\alpha$-lead is defined by $J^{\alpha}_{\sigma}(t) = \frac{d}{dt} \sum_{m} \hat{n}_{\alpha m\sigma}(t) $, where $\hat{n}_{\alpha m\sigma}(t)$ is the spin-resolved density operator for the $\alpha$-lead attached to site $m$.
From the Heisenberg equation of motion, one obtains \cite{Jauho1994,stefanucci2013nonequilibrium}
\begin{dmath}
J^{\alpha}_{\sigma}(t) = 2\Re \sum_{j} [\Sigma^{\alpha}_\sigma * \Gamma_{\sigma}]^<_{jj}(t,t)~,
\end{dmath}
where $\Sigma^{\alpha}_{\sigma}$ is the lead self-energy, see Appendix \ref{sec: lead self-energy}.
The average double occupancy on site $j$ is given by
\begin{dmath}
D_{i}(j) = \sum_{Q} p_j^Q D_i^{Q,\text{imp}}(t)~,
\end{dmath}
where the double occupancy $D_{j}^{Q,\text{imp}}(z) =  \avg{\hat{n}_\uparrow(z) \hat{n}_\downarrow(z)}_{S_{j}^{Q,\text{imp}}}$ for a given local configuration $Q$ can be evaluated from the equation of motion \cite{stefanucci2013nonequilibrium} as
\begin{dmath}
iU^{Q}_j(z) D_j^{Q,\text{imp}}(z) = [\Sigma_{j\sigma}^{Q,\text{imp}} * G_{j\sigma}^{Q,\text{imp}}](z,z^+) + iU_{j}^Q(z) n_{j-\sigma}^{Q,\text{imp}}(z) n_{j\sigma}^{Q,\text{imp}}(z)~.
\end{dmath}
The total energy of the system is the sum of the kinetic and potential energy contributions, 
\begin{dmath}
E^{\text{tot}}(t) = -i\sum_{ij\sigma} W_{ij,\sigma}(t) \Gamma_{ji,\sigma}^<(t,t) 
-i \sum_{j\sigma} \sum_{Q} p_j^Q \left( [\epsilon_{j}^{Q}(z) - \mu ] G_{j\sigma}^{Q,\text{imp}}(z,z^+) + iU_j^Q(z)D_j^{Q,\text{imp}}(z)\right)~.
\end{dmath}
Here, the first term represents the kinetic energy, while the second term corresponds to the local potential energy, which comprises the energies of both singly and doubly occupied states. We furthermore assume in this formula that leads, if present, are represented as additional baths. 

\subsection{Numerical implementation}

The previous formalism is based on the three-branch Kadanoff-Baym contour, which allows to describe general non-equilibrium time evolutions, starting from an initial equilibrium state. With some adaptations, the formalism can also be applied to steady-state situations. This subsection details how time-dependent and steady-state simulations are implemented.

\subsubsection{Generic time-dependent problems}

To solve the equations formulated on the Kadanoff-Baym contour, one could discretize the contour time variables and transform the equations into matrix equations \cite{PhysRevLett.97.266408,PhysRevB.106.195156}.
An alternative is to apply Langreth's rules to transform the contour equations into equations depending on real and imaginary time, and then discretize these variables \cite{RevModPhys.86.779,stefanucci2013nonequilibrium}.
The latter approach has several advantages: 
(i) It preserves the causality of the solution, which ensures that physical quantities at time $t$ are independent of the future evolution of the system.
(ii) It reduces the computational complexity by allowing one to calculate the Green's function from $t=t_0$ step by step, using a small number of iterations at a given time step, rather than trying to converge the solution on the whole contour simultaneously.
(iii) The time-stepping approach makes it easier to implement higher order integration schemes.

Because of causality, the equations on the Matsubara axis form a closed self-consistency loop that can be solved prior to the real-time propagation. 
Physically, this corresponds to preparing the initial equilibrium state.  
Once this initial state has been obtained, one can calculate the real-time (mixed, retarded, and lesser) components of the Green's functions and hybridization functions starting from $t_0$ by incrementing the maximum simulation time step by step. 
Numerical routines for solving the time propagation of the Green's functions with high-order accuracy are implemented in the NESSi library. 
Interested readers are referred to Ref.~\cite{SCHULER2020107484} for more details.

\subsubsection{Steady-state problems}

If the system is coupled to external baths, the initial correlations are expected to be wiped out in the long-time limit \cite{RevModPhys.86.779}. 
This means that the Matsubara (vertical) branch can be neglected in Fig.~\eqref{fig: Kadanoff-Baym contour}, and the three-branch Kadanoff-Baym contour reduces to a two-branch Keldysh contour \cite{Haug2008}.
Only the retarded and lesser real-time components remain in this case.
Furthermore, the restoration of time translational invariance in steady-state situations implies that the Green's functions only depend on the time difference, which enables the use of Fourier transforms and frequency-domain representations. For a general function $f$ we define the Fourier transforms as
\begin{equation*}
f(\omega) = \int_{-\infty}^{\infty} dt f(t) e^{i\omega t}, \quad f(t) = \frac{1}{2\pi}\int_{-\infty}^\infty d\omega f(\omega) e^{-i\omega t}~.
\end{equation*}
In frequency space, the integro-differential Dyson equations as well as the Volterra equations reduce to simple algebraic equations. 
For example, the lattice Dyson equation \eqref{eq: lattice Dyson equation} becomes \cite{Haug2008}
\begin{subequations}
\begin{dmath}
\Gamma^r_{ij,\sigma}(\omega) = \gamma^r_{i\sigma}(\omega) \delta_{ij} 
+ \left[ \gamma^r_{\sigma}(\omega) \tilde{W}_{\sigma}^r(\omega) \Gamma_{\sigma}^r(\omega) \right]_{ij}~,
\end{dmath}
\begin{dmath}
\Gamma_{ij,\sigma}^<(\omega) = \left[ \Gamma_{\sigma}^r(\omega)\tilde{W}^<_{\sigma}(\omega)\Gamma_{\sigma}^a(\omega) \right]_{ij}
+ \sum_m \left(I + \Gamma_\sigma^r(\omega)\tilde{W}_\sigma^r(\omega)\right)_{im} \gamma_{m\sigma}^<(\omega) \\
\mbox{}\hspace{12.3mm}\times\left(I + \tilde{W}_\sigma^a(\omega) \Gamma_\sigma^a(\omega)\right)_{mj}~.
\end{dmath}
\end{subequations}
The equilibrium formalism can be recovered by imposing the fluctuation-dissipation theorem \cite{Haug2008}. 
In the DMFT context, steady-state formalisms have been previously presented in Refs.~\cite{Aron2012,Li2015,*Li2021}.
Here, we use the steady-state framework of Ref.~\cite{PhysRevB.105.085122} and refer the reader to this paper for implementation details.

\section{Numerical results and discussions\label{sec: results}}

\subsection{General remarks}

\begin{figure}
\includegraphics[width=1.0\linewidth]{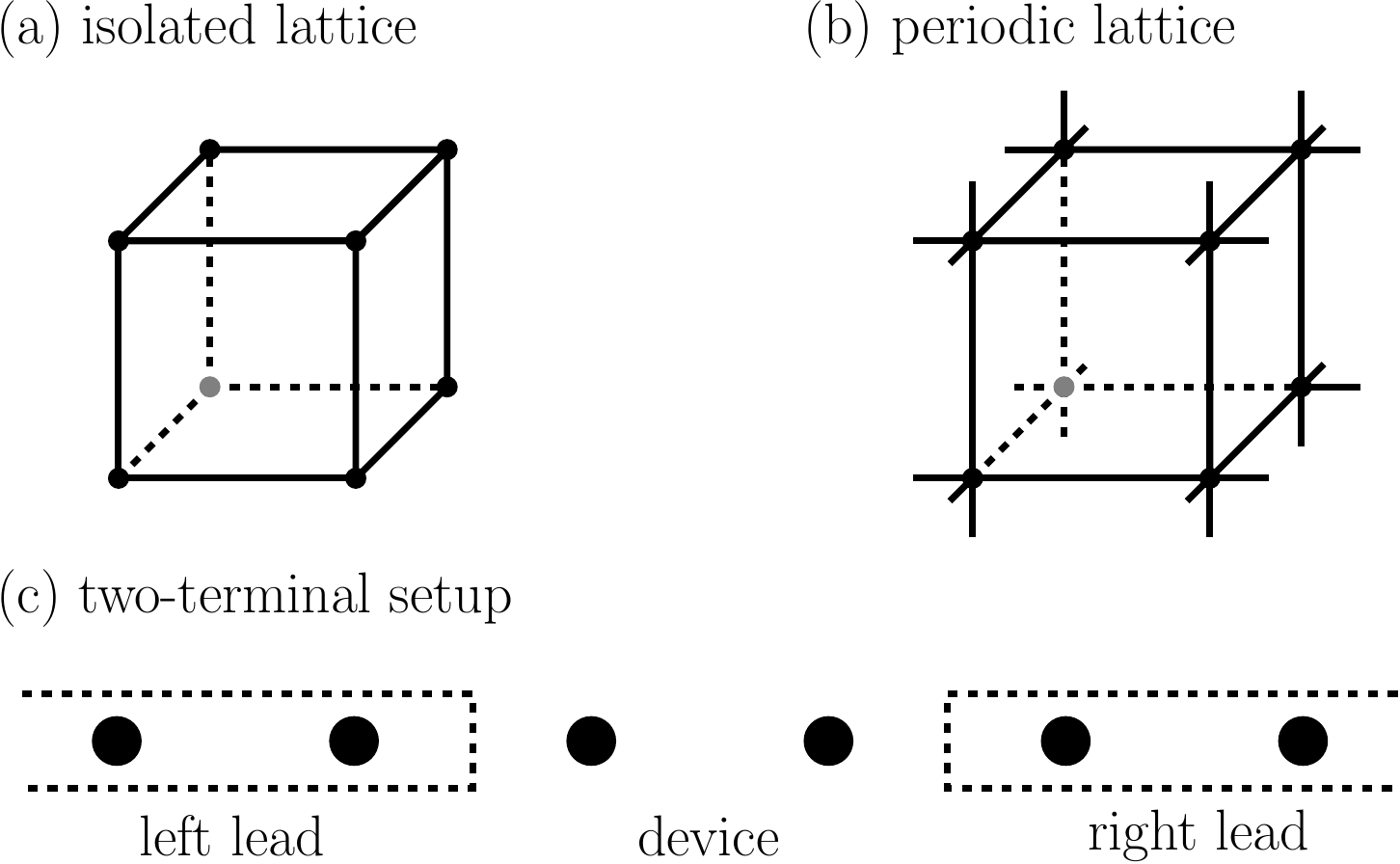}
\caption{Schematic illustration of (a) an isolated cube, (b) a cubic lattice and (c) a two-terminal open structure.
\label{fig: models}}
\end{figure}

In this section, we present the numerical results obtained with our method, focusing on paramagnetic states. We will therefore suppress the spin index $\sigma$ in the following. However, it is worth noting that the method can be easily extended to symmetry broken phases. We discuss three models: (i) a $2$-by-$2$-by-$2$ cube, (ii) a cubic lattice with periodic boundary conditions, and (iii) a small one-dimensional atomic chain, as depicted in Fig.~\ref{fig: models}.
For simplicity, we consider a binary alloy in our calculations, but the formalism is also applicable to multi-component alloys \footnote{ For continuous distributed disorders, one can simulate this by sampling over the distribution function and transforming to the multi-component alloy problem.}.
We refer to the two species of the binary alloy as the host atom and the impurity atom. 

\subsection{$2 \times 2 \times 2$ cubic molecule}

\begin{figure*}
\includegraphics[width=0.9\linewidth]{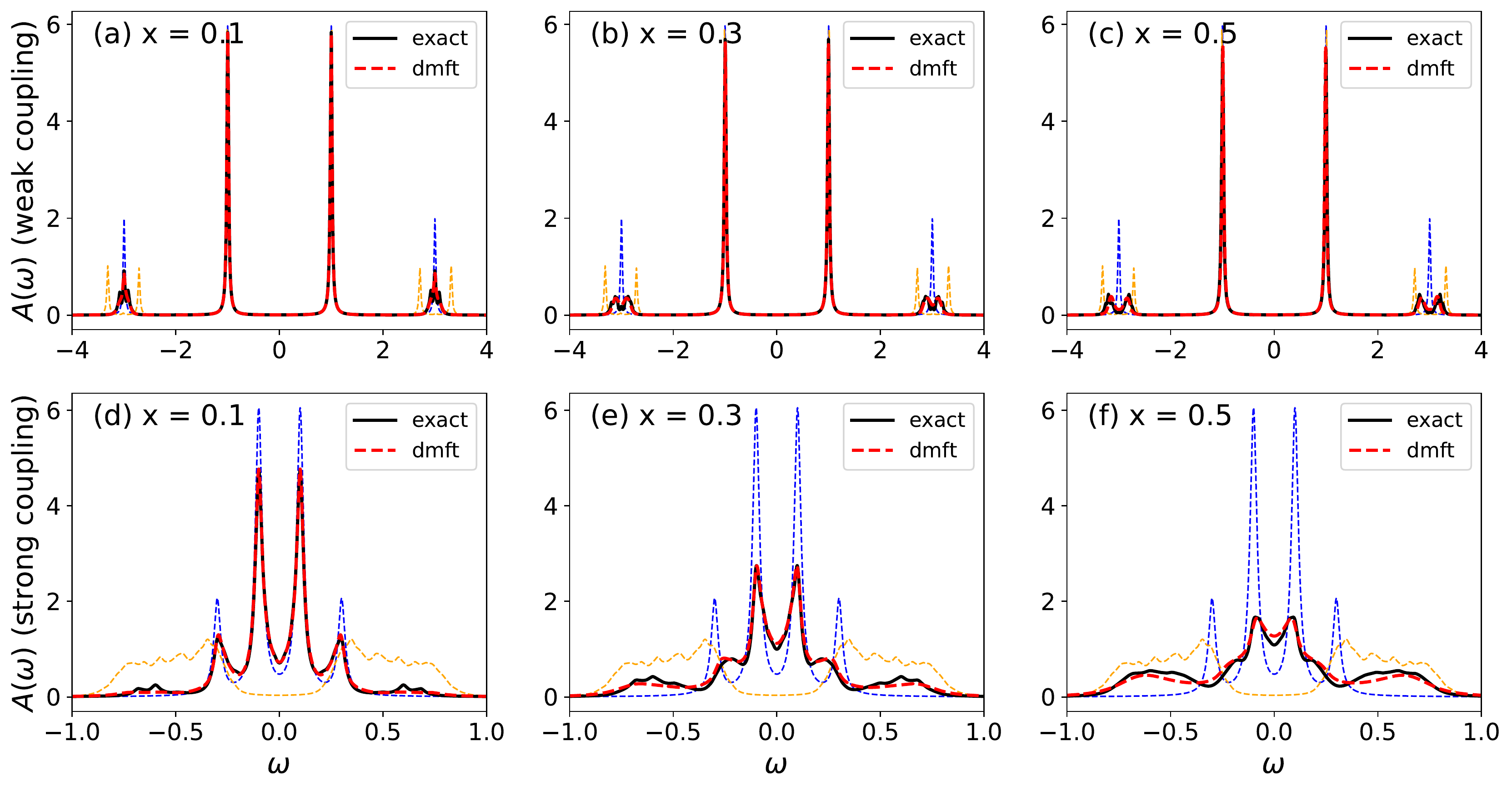}
\caption{
Local spectral function of an $8$-site cube for different impurity concentrations $x$ and coupling strengths (the first and second row is for the weakly and strongly interacting system, respectively). Black solid (red dashed) lines show the exact (DMFT) results, while blue and orange thin dashed lines represent a clean system with only host and impurity atoms, respectively.
\label{fig: site8-spectrum}}
\end{figure*}

To demonstrate the effectiveness of the method, we benchmark it in equilibrium on an isolated $2 \times 2 \times 2$ cube, as shown in Fig.~\ref{fig: models}~(a).
Each site corresponds to a host atom with probability $p_i^{Q=\text{host}} = 1- x$, or an impurity atom with probability $p_i^{Q=\text{imp}} = x$.
The Coulomb and on-site energies of the host and impurity atoms are set to $U_i^{\text{host}} = \epsilon_i^{\text{host}} = 0$ (non-interacting) and $U_i^{\text{imp}} = -2\epsilon_i^{\text{imp}} = 1$, respectively \footnote{The parameters are chosen to be appropriate for half-filling, since IPT solver gives reasonable results in this regime.}.
Only nearest-neighbor hopping is considered, with a value of $W_{\langle ij\rangle} = 1$ in the weak coupling case and $W_{\langle ij\rangle} = 0.1$ in the strong coupling case.
The exact solution can be calculated by diagonalizing the many-body Hamiltonian in the Hilbert space with dimension $4^{8} = 65536$ and averaging over the configurational space with dimension $2^8 = 256$.
We employ an inverse temperature of $\beta = 20$ and a broadening parameter $\eta=0.01$ when plotting the spectral functions $A(\omega) = -\frac{1}{\pi}\Im G^r(\omega)$, which are obtained from the retarded component of the Green's functions.
The retarded Green's function can be calculated using the Lehmann representation, once we know the many-body states $\ket{n}$ and the corresponding eigenvalues $\omega_n$ for a given configuration, 
\begin{dmath}\label{eq: Lehmann representation}
G^r_{ij,\sigma}(\omega) = \frac{1}{Z}\sum_{mn}\left(e^{-\beta \omega_n} + e^{-\beta \omega_m}\right)\frac{\avg{n|c_{i\sigma}|m}\avg{m|c_{j\sigma}^\dagger|n}}{\omega + \omega_n - \omega_m + i\eta}~,
\end{dmath}
where $Z = \sum_n e^{-\beta \omega_n}$ is the partition function.

Figure~\ref{fig: site8-spectrum} shows the (disorder averaged) local spectral functions in the weakly (first row) and strongly (second row) interacting systems, for the indicated impurity concentrations $x$.
The black and red lines refer to the exact and DMFT results, respectively. 
In addition, the spectrum of the clean system with only host (impurity) atoms is displayed using blue (orange) thin dashed lines.

In the weak coupling case with $W_{\langle ij\rangle} = 1$, the non-interacting spectral function, represented by the blue dashed line in Fig.~\ref{fig: site8-spectrum}~(a-c), shows four peaks at $\omega = \pm 3\text{, }\pm 1$ due to the bonding and anti-bonding states produced by the hopping between the sites.
As we increase $U$ from $0$ to $1$ in the homogeneous system, the non-interacting peaks at $\omega = \pm 3$ split into two peaks, as shown by the orange dashed lines.
The exact spectral functions of the disordered systems, shown by the black solid lines, are in between these two spectra.
With increasing $x$, the peaks at $\omega= \pm 3$ split, but the separation between the subpeaks is smaller than in the uniform interacting system.
The red dashed line plots the DMFT results, which agree very nicely with the exact results for all impurity concentrations.

In the case of strong coupling, i.e. for $W_{\langle ij\rangle} = 0.1$, the bonding and anti-bonding states of the non-interacting system (represented by blue dashed lines) are located at $\omega = \pm 0.1$ and $\omega = \pm 0.3$, respectively (see Fig.~\ref{fig: site8-spectrum}~(d-f)).
When $U=1$, two Hubbard bands appear at approximately $\omega = \pm 0.5$, as depicted by the orange dashed lines. 
With increasing impurity concentration, spectral weight from the bonding and anti-bonding peaks is transferred to the Hubbard bands, resulting in a complicated spectral structure for large $x$. 
Nevertheless, the DMFT results still agree nicely with the exact results, demonstrating the effectiveness of the method for the description of this small-size system with coordination number $z=3$.

\begin{figure}
\includegraphics[width=1.0\linewidth]{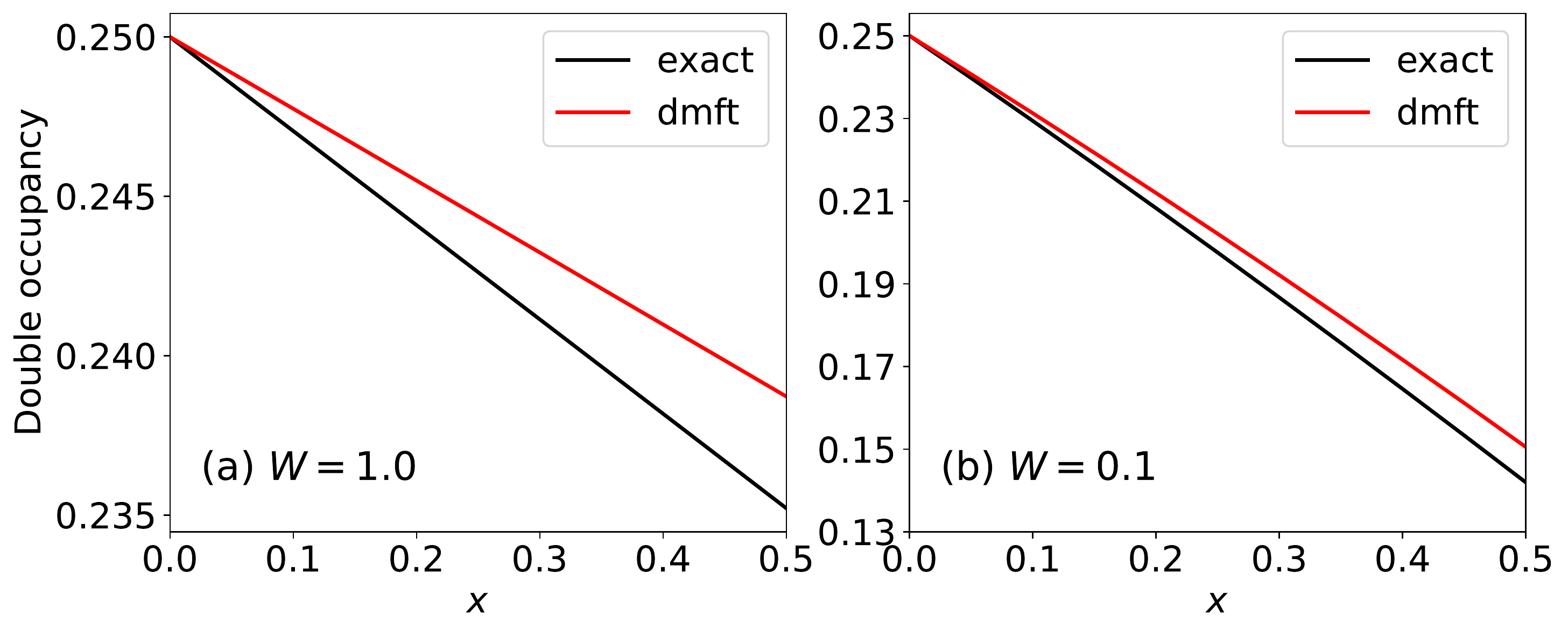}
\caption{Double occupancy as a function of impurity concentration $x$ for the $8$-site cube in the case of (a) weak coupling ($W_{\avg{ij}} = 1$) and (b) strong coupling ($W_{\avg{ij}} = 0.1$).
Black lines show the exact results and red lines the approximate DMFT results.
\label{fig: site8-docc}}
\end{figure}

Figure~\ref{fig: site8-docc} shows the double occupancy as a function of $x$ for (a) weak coupling and (b) strong coupling, with the exact and DMFT results represented by black and red lines, respectively.
As expected, the double occupancy decreases with increasing $x$ due to the Coulomb interaction on the impurity atoms.
It is worth noting that the DMFT results slightly overestimate the double occupancy in both cases.
There are two main factors responsible for the discrepancy between the exact and DMFT results:
(i) In DMFT, the self-energy of the lattice system is approximated to be site-diagonal. This assumption neglects non-local correlations, which are relevant in finite-connectivity systems.
(ii) Additionally, the IPT impurity solver considers only a finite set of selected diagrams for the self-energy of the single impurity Anderson model, which can introduce a bias and result in an inaccurate description of local time-dependent fluctuations.

\subsection{Cubic lattice}

\begin{figure}
\includegraphics[width=1.0\linewidth]{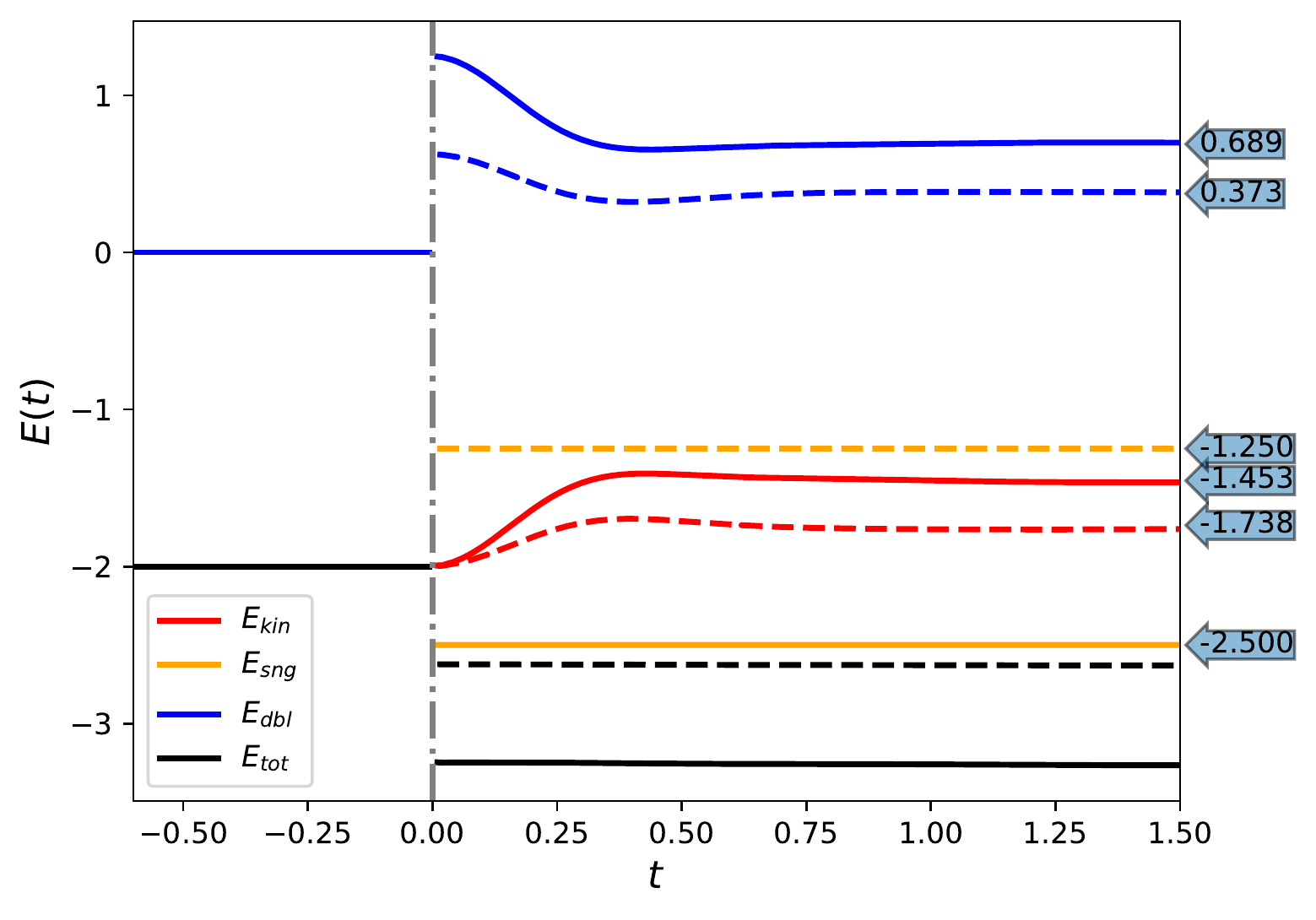}
\caption{Energies as a function of time for a quench from $U_i = -2\epsilon_i = 0 $ to $U_i = -2\epsilon_i = 5$ at $t=0$ in a half-filled system. Red, orange, blue and black lines represent the kinetic energy, singly occupied potential and doubly occupied potential energies, and total energy, respectively. 
The solid (dashed) lines correspond to the case where all (half) of the sites undergoing the quench.
The arrows indicate the values of the thermalized system. 
\label{fig: cubic-quench}}
\end{figure}

We next study an interaction quench problem for a three-dimensional cubic lattice, as shown in the inset of Fig.~\ref{fig: cubic-quench}.
The nearest-neighbor hopping $W_{\langle ij\rangle} = 1$ serves as the energy unit and $\hbar/W_{\langle ij\rangle}$ as the unit of time.
The non-interacting system is initially prepared in an equilibrium state with inverse temperature $\beta = 10$. At $t = 0$, we apply a quench, which suddenly changes $U_i = -2\epsilon_i = 0 $ to $U_i = -2\epsilon_i = 5$ on all sites (a disordered system with $50\%$ sites undergoing a quench is discussed later).
For this choice of parameters the system exhibits particle-hole symmetry, which ensures that the electron density per spin equals $0.5$ during the whole time evolution. 
The red, orange and blue solid lines in Fig.~\ref{fig: cubic-quench} show the time evolution of the corresponding kinetic energy $E_{\text{kin}}$, the singly-occupied contribution to the potential energy $E_{\text{sng}}$, and the doubly-occupied contribution to the potential energy $E_{\text{dbl}}$, respectively. 
The total energy $E_{\text{tot}}$, which is the sum over these three contributions, is represented by black solid lines.
The right arrows in Fig.~\ref{fig: cubic-quench} indicate the energy values of an equilibrium system with an effective temperature, as discussed below.

In the initial noninteracting state, both the singly occupied ($E_{\text{sng}}$) and doubly occupied ($E_{\text{dbl}}$) potential energies are zero, while the kinetic energy ($E_{\text{kin}}$) is $-2$, resulting in a total energy $E_{\text{tot}} = -2$, as shown in Fig.~\ref{fig: cubic-quench}. During the quench at $t=0$, $E_{\text{sng}}$ and $E_{\text{dbl}}$ abruptly change from $0$ to $2\epsilon_i n_i = -2.5$ and $Un_i^2 = 1.25$, respectively, because of the sudden modification of $\epsilon_i$ and $U_i$. In contrast, the evolution of the kinetic energy $E_{\text{kin}}$ is continuous, since the hopping integral $W_{\avg{ij}}$ does not experience a quench.
Within approximately one inverse hopping time after the quench, the system appears to be thermalized.
$E_{\text{sng}}$ remains constant during this process due to the constant electron density ($n_i(t) = 0.5$ for each spin channel), which is protected by the particular form of the quench.
However, the kinetic energy of the electrons increases while the potential energy contribution associated with the double occupancy decreases during this period.
The loss of potential energy compensates the gain in kinetic energy, resulting in a constant total energy (black line).
This is an expected consequence of energy conservation, since the system is isolated during the time evolution and there is no energy exchange with an environment.
It is worth noting that although the bare IPT impurity solver used in these calculations is not conserving in the Baym-Kadanoff sense \cite{PhysRev.127.1391,*PhysRev.124.287}, and energy is hence not exactly conserved, it produces an almost constant total energy if the interaction after the quench is not too strong \cite{Tsuji2013}.

We next study a disordered situation, where only half the sites undergo the quench, using the same parameters as in the previous calculation. 
(The other sites remain noninteracting.) 
The evolution of the various energy contributions in this system is represented by the dashed lines in Fig.~\ref{fig: cubic-quench}, with the same color scheme as before.
Although the quench induced changes are qualitatively similar to those observed in the clean system, the amplitude of the change becomes weaker due to the interpolation between the non-interacting and uniformly quenched solutions. 
This is to be expected in a disordered system, and the results support the validity of the DMFT approach.
In particular, we note again that the total energy remains essentially constant, which indicates that for the present parameters, DMFT treats the different energy contributions in a thermodynamically consistent way even in the presence of disorder.

\begin{figure}
\includegraphics[width=1.0\linewidth]{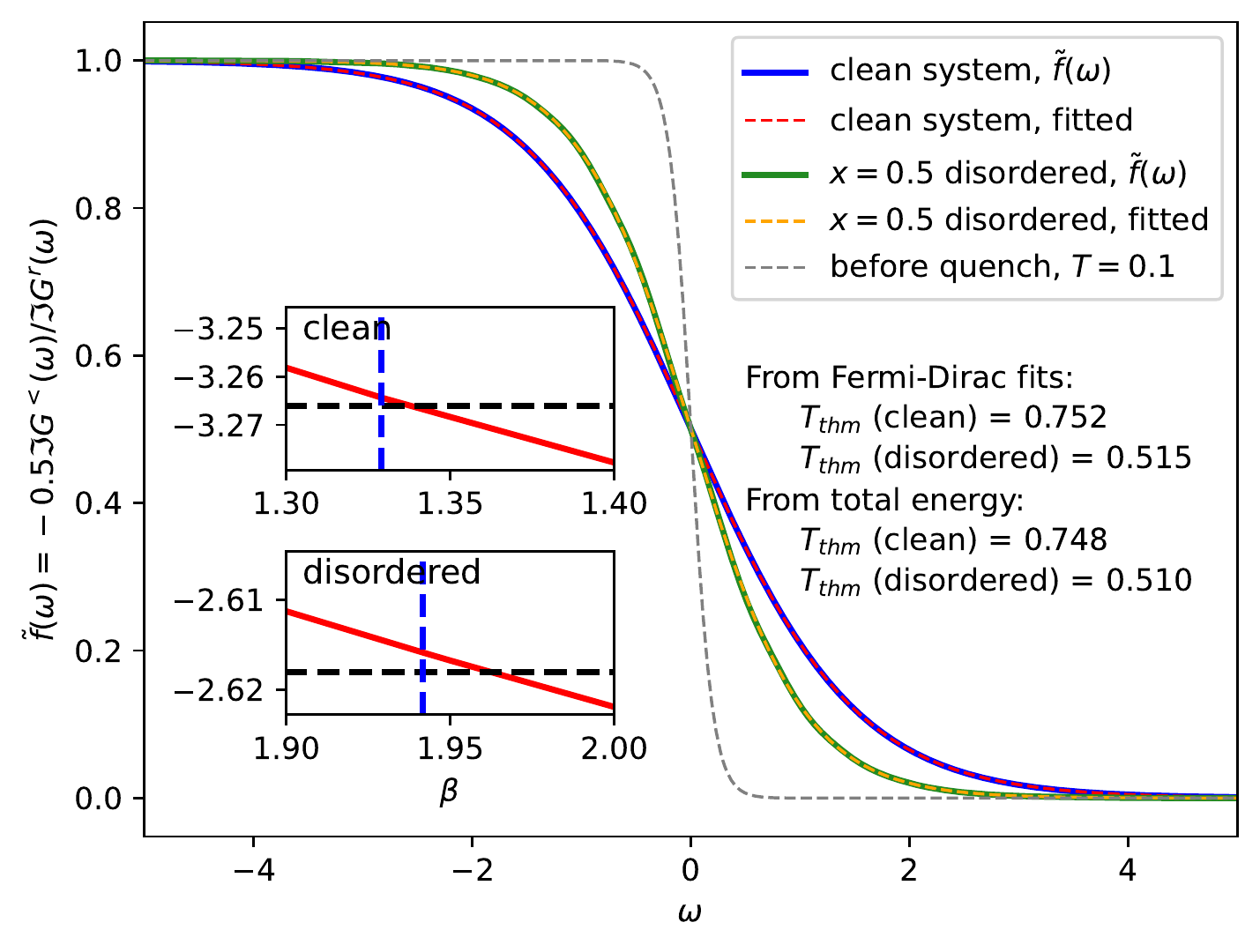}
\caption{Onsite nonequilibrium distribution function $\tilde{f}(\omega, t_{\text{av}} = 5)$ for clean (blue) and disordered (green) systems. Red and orange dashed lines are fitted Fermi-Dirac functions with temperature $T_{\text{thm}}=0.752$ (clean system) and $T_{\text{thm}}=0.515$ (disordered system), respectively. The grey dashed line shows the Fermi-Dirac function for the initial temperature $T = 0.1$.
Insets: The red line shows $E_{\text{tot}}$ as a function of inverse temperature for equilibrium interacting systems. The dashed horizontal line plots $E_{\text{tot}}$ after the quench.
The dashed vertical line gives the temperature from Fermi-Dirac fits.
\label{fig: cubic-quench-dist}}
\end{figure}

We finally investigate the thermalization process following the quench. To analyze this, we introduce an onsite non-equilibrium distribution function,
\begin{dmath}\label{eq: nonequilibrium distribution function}
\tilde{f}_i(\omega, t_{\text{av}}) = -\frac{1}{2} \frac{\Im G_i^<(\omega, t_{\text{av}})}{\Im G_i^r(\omega, t_{\text{av}})}~.
\end{dmath}
Here, $G(\omega,t_{\text{av}}) = \int_{-\infty}^\infty d t_{\text{rel}} e^{i\omega t_{\text{rel}}} G(t,t')$ with $t_{\text{av}}=(t+t')/2$ and $t_{\text{rel}}=t-t'$ is the Wigner representation of the two-time Green's function \cite{RevModPhys.86.779}. 
$\tilde{f}$ reduces to the Fermi-Dirac distribution function in equilibrium, where the fluctuation-dissipation theorem holds \cite{Haug2008}.
In Fig.~\ref{fig: cubic-quench-dist}, the blue and green solid lines correspond to $\tilde{f}_i(\omega, t_{\text{av}} = 5)$ for the quench of the clean and disordered systems, respectively. Recall that, for the disordered case, half of the sites undergo the quench. 
We fit the curves with Fermi-Dirac functions, as shown by the red and orange dashed lines. 
As a reference, we also plot the distribution function before the quench ($T=0.1$) as the grey line. It can be observed that the Fermi-Dirac function fits well in both cases, indicating complete thermalization after the quench due to electron-electron scattering. 
The fits yield the temperatures $T_{\text{thm}} = 0.752$ (clean) and $T_{\text{thm}} = 0.515$ (with disorders) of the thermalized systems. 
We can now calculate the kinetic and potential energies of the equilibrium systems with $T_{\text{thm}}$, which nicely match with the values of the quenched systems at times $t\gtrsim 1$, as illustrated by the arrows in Fig.~\ref{fig: cubic-quench}.

To check if the system is really thermalized, we determine the fully thermalized temperature from the total energy. For this we determine the temperature of equilibrium systems with the post-quench parameters, such that the total energy matches the total energy after the quench.
Specifically, for the clean system, we measure (at $t=1.5$) a total energy of $E_{\text{tot}} = -3.266$, and for the disordered system $E_{\text{tot}} = -2.618$, which are plotted as dashed horizontal lines in the insets of Fig.~\ref{fig: cubic-quench-dist}. The red lines in the same insets show the temperature dependence of the total energy in equilibrium.  The intersects of the red solid and dashed horizontal lines determine the (inverse) temperatures of the fully thermalized systems. 
In particular, $T'_{\text{thm}} = 0.748$ for clean system and $T'_{\text{thm}} = 0.510$ for disordered system.
The blue dashed vertical lines indicate the effective (inverse) temperatures from the Fermi function fits, and one can see that these temperatures are close to the fully thermalized values, both in the clean and disordered systems. The small discrepancies may be due to the fact that the systems at $t=5$ are not yet completely thermal, or they could be a consequence of the fact that the bare IPT solver does not fully conserve the total energy.   

\subsection{Atomic chain}

In the third example, we study a short atomic chain consisting of two central scattering sites (generically with disorder and Coulomb interactions) sandwiched between two leads, as depicted in Fig.~\ref{fig: models}~(c). The leads are assumed to be non-interacting and without disorders.
The two central scattering sites of our system will be referred to as the left and right (scattering) sites in the following. 

\subsubsection{Non-interacting chain: CPA treatment}

In the first investigation, we use the same parameters as in Fig.~5~(a) of Ref.~\onlinecite{PhysRevB.94.075426}. 
Specifically, we choose the hopping amplitude between the left and right sites, $W$, as the energy unit ($W=1$). 
The onsite energies for the host and impurity atoms in the central device region are $0$ and $0.5$, respectively, and the impurity concentration is $x=0.3$. 
Consistent with Ref.~\onlinecite{PhysRevB.94.075426}, we employ the wide-band limit (WBL) for the leads, a coupling strength of $0.5$ for both the left and right leads, and inverse lead temperature $\beta = 10$.
After $t = 0$, constant voltages of $V_L=1.5$ and $V_R=-1.5$ are applied to the left and right electrodes, respectively, by uniformly shifting the on-site energies of the leads.
This results in an electron flow from the left lead to the right lead.
If the central lattice sites have no interactions, our formalism reduces to the time-dependent CPA.

\begin{figure}[t]
\includegraphics[width=1.0\linewidth]{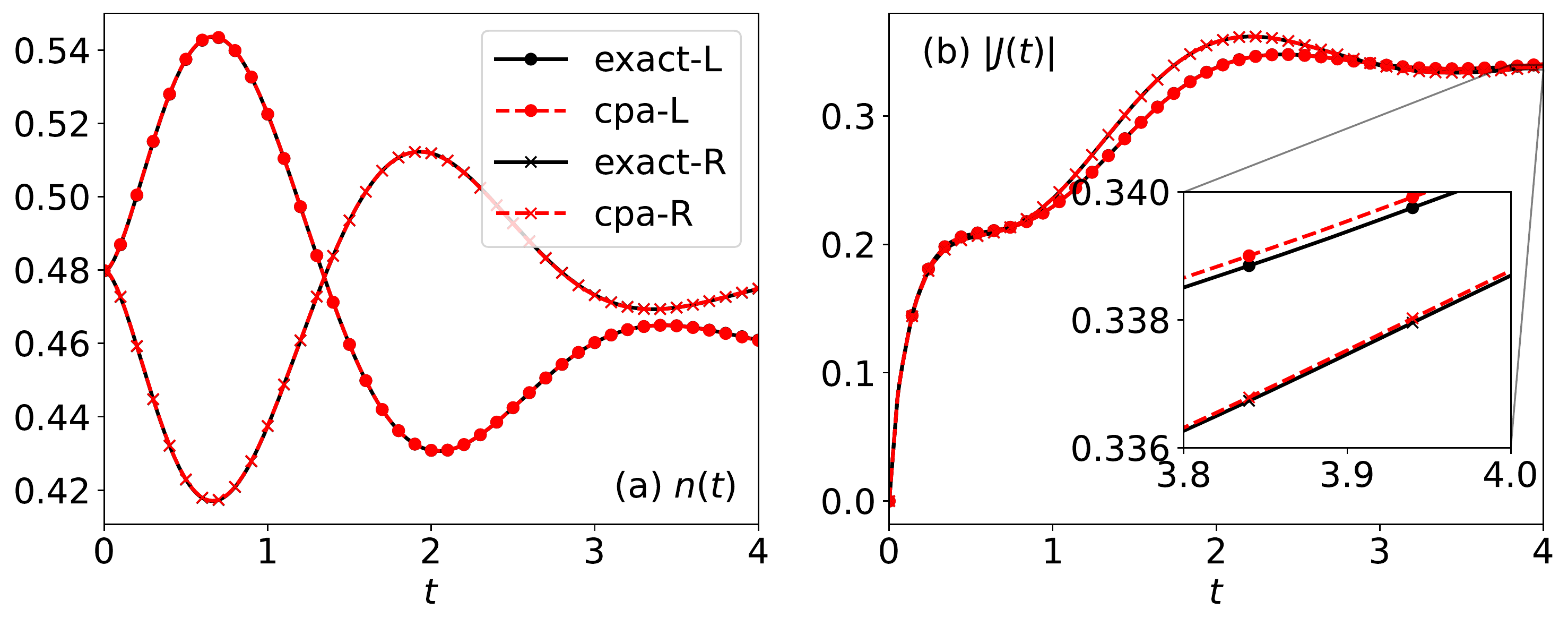}
\caption{Evolution of (a) the electron density and (b) the current for a non-interacting disordered $2$-site chain after a step-shaped voltage pulse applied to the leads. Black solid and red dashed lines show the exact and the CPA results, respectively. Lines with circles are for the left site and lines with crosses for the right site. 
\label{fig: atomic chain: density and current}}
\end{figure}

Figure~\ref{fig: atomic chain: density and current} (a) plots the electron densities on the left (circle marker) and right (cross marker) sites as a function of time, obtained from both exact  (black solid line) and CPA (red dashed line) calculations.
For the exact reference, we average the results obtained with NESSi for the four possible impurity configurations.  
After the voltage quench, the electron density on the left site starts to increase, while that on the right site decreases due to the flow of electrons from the left lead into the structure, and from the structure to the right lead.
The electron densities exhibit damped oscillations, and are expected to reach some steady-state values at longer times. 
The asymmetry in the densities on the left (circle marker) and right (cross marker) sites arises from the impurity sites, which drive the system away from half-filling (the impurity onsite energy is $0.5$).

Figure~\ref{fig: atomic chain: density and current}~(b) shows the absolute value of the time-dependent current flowing out of (into) the left (right) lead. The inset provides a zoomed-in view of the latest times.
The current approaches a nonzero steady-state value after several oscillations.
Note that in the transient regime, the currents running through the left and right leads are, in general, not equal, since there can be charge accumulation in the central sites.
However, they approach the same steady-state values once the occupations of the sites have settled to their steady-state values.
One can see that the CPA results agree remarkably well with the exact data for both the electron densities and the currents in the parameter regime considered in these calculations.
Additionally, Fig.~\ref{fig: atomic chain: density and current}~(b) agrees well with Fig.~5~(a) in Ref.~\onlinecite{PhysRevB.94.075426}, where the authors developed a time-dependent CPA with nonequilibrium vertex corrections on a two-branch Keldysh contour \footnote{Our calculation uses an inverse temperature of $\beta = 0.05$ instead of zero temperature as in Ref.~\onlinecite{PhysRevB.94.075426}, but this does not significantly affect the results.}.
We would like to point out, however, that while the WBL is required in the formalism of Ref.~\onlinecite{PhysRevB.94.075426}, this is not necessary here, since any leads (possibly with real dispersion relations) can be implemented in principle.

\subsubsection{Generic interacting disordered chain}

\begin{figure}
\includegraphics[width=1.0\linewidth]{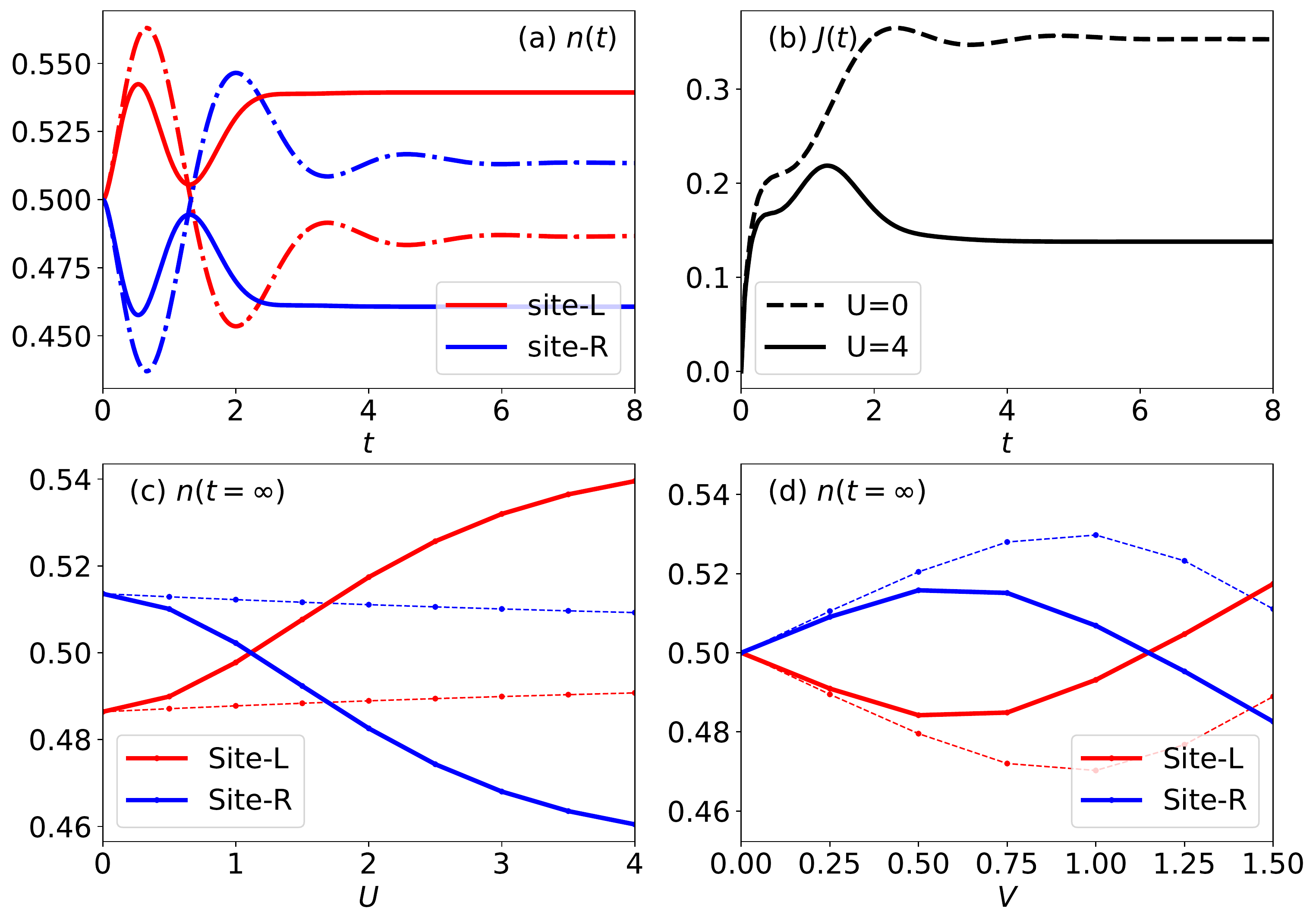}
\caption{Time dependent electron density (a) and current (b) for an interacting 2-site model coupled to two leads subject to a step-shaped voltage pulse. 
Dashed lines are for the non-interacting case and solid lines for $U=4$.
The lower panels show the steady-state electron density $n(t=\infty)$ versus $U$ with fixed $V=1.5$ (c), and $n(t=\infty)$ versus $V$ with fixed $U=2$ (d). 
Thin dashed lines and bold solid lines show the Hartree and DMFT solutions, respectively.
\label{fig: chain_clean}}
\end{figure}

We now turn our attention to a case with interacting electrons in the central scattering region. 
To maintain the system close to half-filling, we keep the parameters for the leads unchanged and set $\epsilon_i = -U_i / 2$. 
With this, the IPT impurity solver provides reasonable results \cite{Tsuji2013}.

Our initial focus is on the clean system, with identical interactions on both sites. Fig.~\ref{fig: chain_clean}~(a) plots the charge density as a function of time for both the left site (red line) and right site (blue line). 
The dashed and solid lines in Fig.~\ref{fig: chain_clean}~(a) correspond to the non-interacting ($U=0$) and interacting ($U=4$) systems, respectively.
Because of the symmetric set-up, the deviations of the charge densities on the left and right sites from their half-filled values $0.5$ are symmetric, i.e. $n_{L}(t) + n_{R}(t) = 1$.
Fig.~\ref{fig: chain_clean}~(b) plots the current, with the red dashed line and black solid line corresponding to the systems with $U=0$ and $U=4$, respectively.
It should be noted that the absolute value of the current out of the left and into the right leads is the same because of the particle-hole symmetric parameters used in the calculation.
Compared to the $U=0$ result, the current is suppressed when $U=4$, since the onsite repulsion creates a large splitting between the local many-body states (Coulomb blockade effect). 
In both the non-interacting and $U=4$ cases, the system exhibits a transient regime before reaching some steady-state value for the current.
Coulomb interactions dampen the oscillations, so that the interacting system approaches the steady-state faster than the noninteracting one. 
Furthermore, we observe that the steady-state distribution of the electrons in the central region can be reversed with increasing $U$, as illustrated in Fig.~\ref{fig: chain_clean}~(c). 
Specifically, for $U=0$, we have $n_L(\infty) < n_R(\infty)$, while for $U=4$, we obtain $n_L(\infty) > n_R(\infty)$.
This is because in the non-interacting case, the rate of electrons (proportional to hopping integral) transferred from the left site to the right site is higher than the rate at which electrons transfer from the right site to the right lead, which leads to charge accumulation on the right site \cite{Datta2005}. 
(If the hopping between the central sites were smaller than between the leads and the central region, the result would be opposite.)
In the system with $U=4$, the Coulomb interaction suppresses the hopping between the central sites, which leads to charge accumulation on the left site.  

We further investigate this effect by plotting $n_{L/R}(\infty)$, i.e., the steady state occupation, as a function of $U$ for fixed $V=1.5$ in Fig.~\ref{fig: chain_clean}~(c), and as a function of $V$ for fixed $U=2$ in Fig.~\ref{fig: chain_clean}~(d).
The solid lines and dashed lines correspond to the DMFT and Hartree results, respectively.
As shown in panel (c), with fixed bias $V=1.5$ and for small $U$, we find $n_L(\infty) < n_R(\infty)$.
However, as $U$ is increased, the curves intersect at some point, and beyond this point, $n_L(\infty) > n_R(\infty)$.
The comparison with the Hartree solution, which does not exhibit this crossing, suggests that this reversal originates from higher-order interaction effects. 
In Fig.~\ref{fig: chain_clean}~(d), where the interaction is fixed to $U=2$ and the voltage $V$ is varied from $0$ to $1.5$, we observe that for low bias $n_L(\infty) < n_R(\infty)$ and the occupation of the left (right) site initially decreases (increases), reaches a minimum (maximum) value, and then starts to increase (decrease).
The Hartree solution shows the qualitatively same behavior as DMFT; however, DMFT shift the crossing point to much lower voltages, compared to the Hartree result.
In both situations, the occupations $n_L(\infty)$ and $n_R(\infty)$ reflect the trade-off between the electrons' ease of hopping to the lead or to another site.
For the same reason, $n_L(\infty)$ and $n_R(\infty)$ also show a crossing in the non-interacting case when decreasing $W$ from $1$ to a small value (not shown).

\begin{figure}
\includegraphics[width=1.0\linewidth]{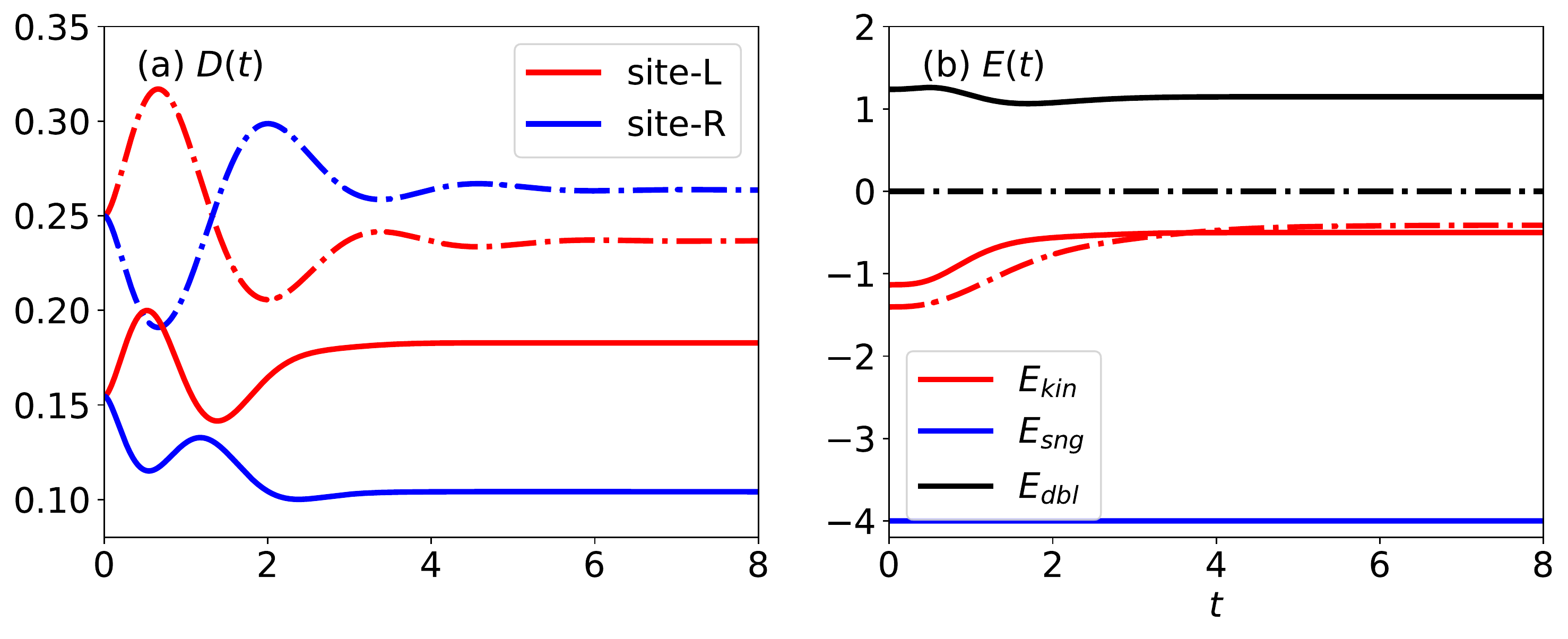}
\caption{
(a) Double occupancy and (b) kinetic and potential energies versus time.
Dashed lines are for $U=0$ and solid lines for $U=4$.
\label{fig: chain_clean_docc}}
\end{figure}

In addition, we also investigate the time-dependent double occupancy $D(t)$ and the evolution of various energy components, as shown in Fig.~\ref{fig: chain_clean_docc}~(a) and (b). 
If $U=0$, $D_{L/R}(t)$ is equal to $n_{L/R}^2(t)$ since electrons do not interact with each other. 
For $U=4$, the double occupation is suppressed due to the Coulomb energy. It is worth noting that $D_{L/R}(t)$ does not exhibit a mirror symmetry around the initial value, even in the non-interacting case. 
Fig.~\ref{fig: chain_clean_docc}~(b) shows the kinetic energy $E_{\text{kin}}$ (red), singly-occupied potential energy $E_{\text{sng}}$ (blue), and doubly-occupied potential energy $E_{\text{dbl}}$ (black), respectively.
$E_{\text{kin}}$, which accounts only for the inter-site hopping between the two central scattering sites, increases after switching on the voltage, 
since the electron distribution becomes nonthermal. In lattice systems, the kinetic energy can be expressed as $E_{\text{kin}}=\sum_k\epsilon_k n_k=\int d\omega \rho(\omega) \omega n(\omega)$, with $\rho$ the density of states. Hence, if the occupation of the electrons becomes flatter (``heating") or high energy states get populated (``inversion"), the kinetic energy increases. 
$E_{\text{sng}}$ remains constant since the total electron number $n(t) = n_L(t) + n_R(t)$ is constant in our symmetric setup, as discussed previously. 
$E_{\text{dbl}}$ exhibits small variations for $U=4$ since the contributions from one site are largely compensated by the other site, as can be seen from panel (a). ($E_{\text{dbl}}$ is proportional to $D(t)$.) 
Note that the total energy in this open setup is not conserved.

\begin{figure}
\includegraphics[width=1.0\linewidth]{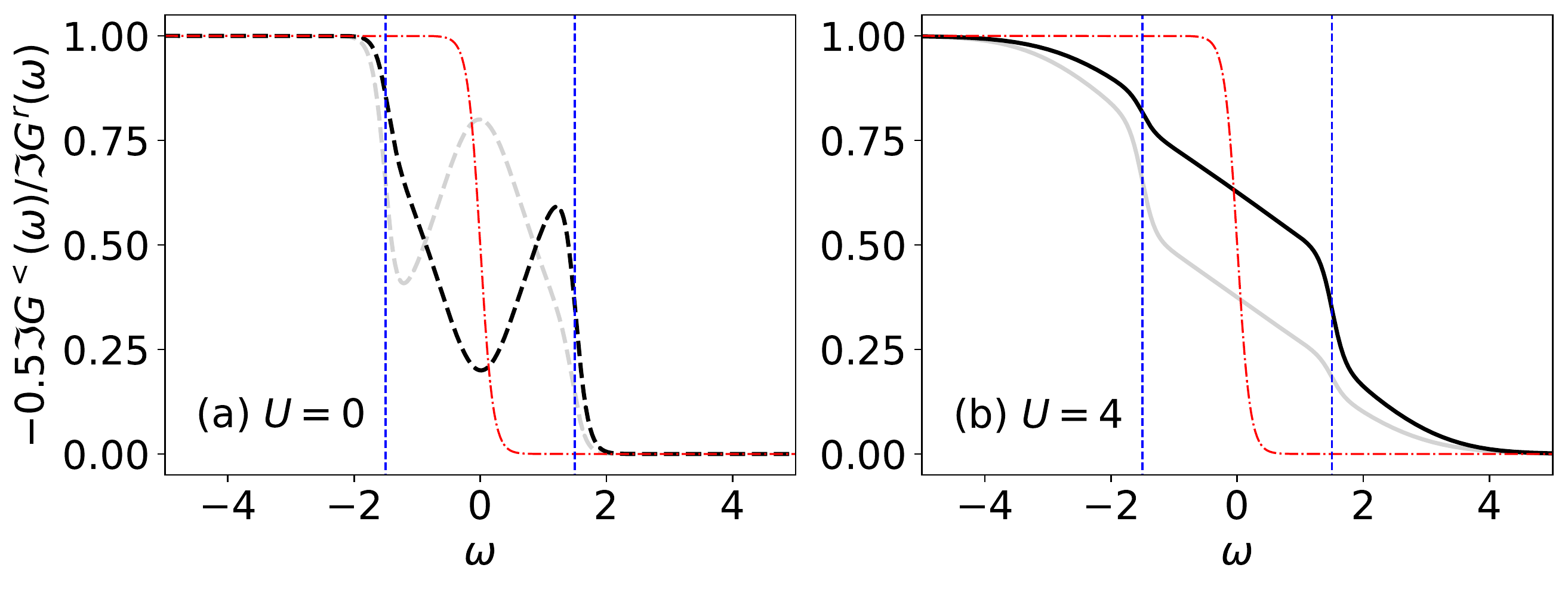}
\caption{Nonequilibrium steady-state distribution function $\tilde{f}_i(\omega) = -0.5 \Im G_i^<(\omega) / \Im G_i^r(\omega)$ with $V=1.5$ for (a) the non-interacting and (b) the interacting ($U=4$) systems. Black (grey) lines show the distributions for the left (right) dots. Blue vertical lines indicate the local chemical potential of the leads. Red dashed lines show the Fermi-Dirac distribution at equilibrium.
\label{fig: chain_heatup}}
\end{figure}

To analyze the nonthermal electron distribution, we plot the steady-state non-equilibrium distribution function $\tilde{f}_i(\omega) = -\frac{1}{2} \frac{\Im G_i^<(\omega)}{\Im G_i^r(\omega)}$
in the presence of a voltage bias $V=1.5$ for $U=0$ and $U=4$ in Fig.~\ref{fig: chain_heatup}~(a) and (b), respectively.
The black and grey lines show the results for the left and right sites.
Blue vertical lines indicate the local chemical potentials of the leads, while the red dashed line corresponds to the Fermi-Dirac distribution with inverse temperature $\beta=10$.
One can see that for $U=0$, $\tilde{f}_i(\omega)$ exhibits a partial population inversion, which is quite different from the superposition of two Fermi-Dirac distributions. 
Specifically, in panel (a), $\tilde{f}_i(\omega)$ on the left site exhibits a hump below $\omega=2$ due to the injection of electrons from the left lead. 
However, this hump is smeared out in the presence of el-el interactions (panel (b)), which help to redistribute the electron population and bring the system into a state with an approximately defined high electronic temperature.

\begin{figure}
\includegraphics[width=0.9\linewidth]{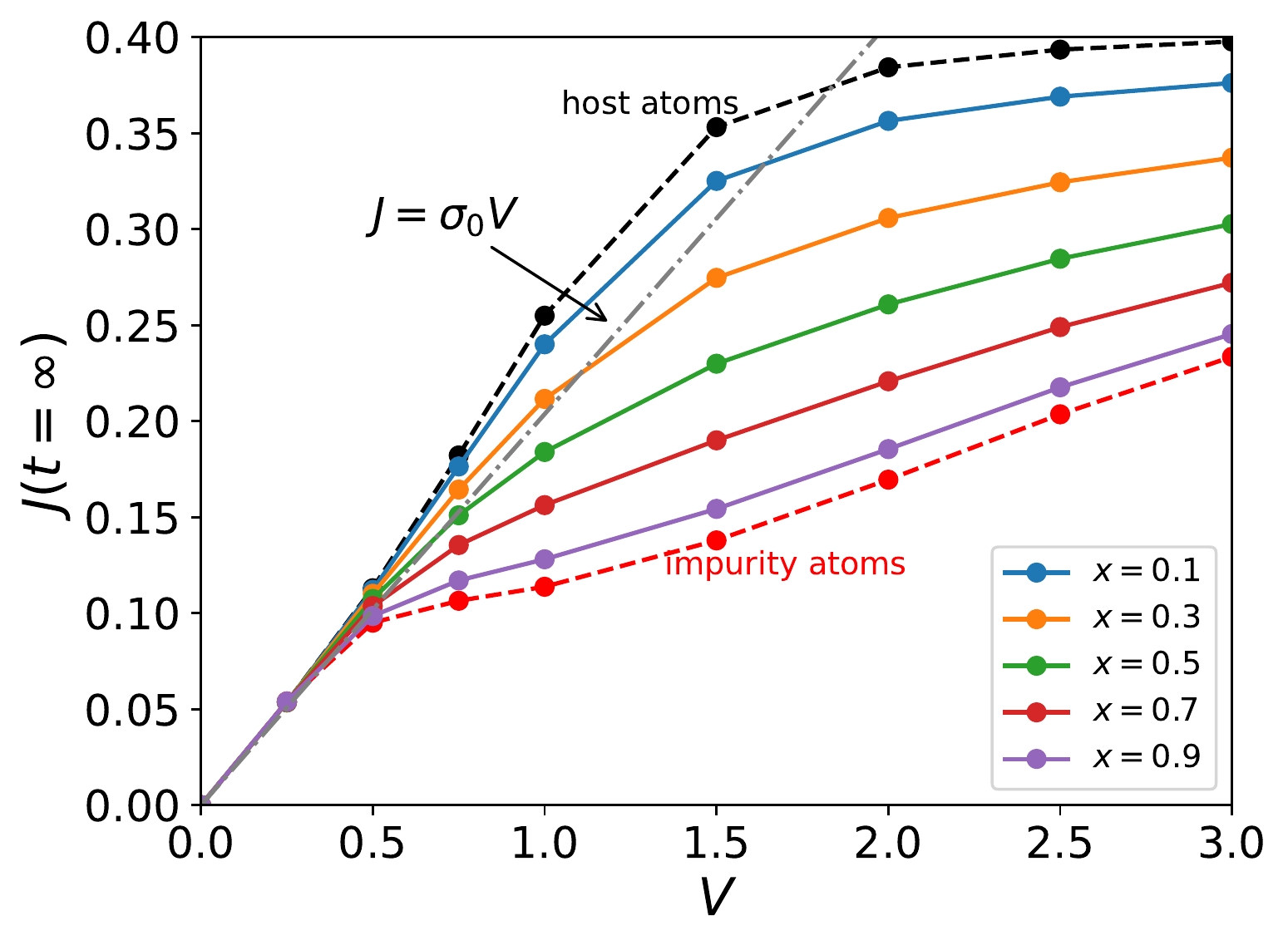}
\caption{Current-voltage characteristics  with different impurity concentrations of the 2-site atomic chain depicted in Fig.~\ref{fig: models}~(c).
The black and red dashed lines are for the host ($U=0$) and impurity ($U=4$) atoms without disorders.
\label{fig: chain_IV}}
\end{figure}

Finally, we investigate the effect of disorder on the current-voltage (IV) characteristics.
The host and impurity parameters are chosen as $U_{\text{host}} = -2\epsilon_{\text{host}} = 0$ and $U_{\text{imp}} = -2\epsilon_{\text{imp}} = 4$, while the lead parameters are kept the same. 
Fig.~\ref{fig: chain_IV} shows the IV characteristics in the steady-state for various impurity concentrations $x$.
In the non-interacting system (black dashed line), with increasing voltage bias $V$, the current initially increases with a slope corresponding to the zero-bias conductance $\sigma_0$ ($\sigma_0 = 0.64/\pi$ for $U=0$ in our case), as shown by the grey dotted line.
At some larger $V$, determined by the finite width of the DOS of the central sites (see below), the current saturates.
In the interacting case, the steady-state current is suppressed (red dashed curve) and a plateau-like structure forms at intermediate $V$, in qualitative agreement with previous quantum dot calculations, see for example Fig.~10 in Ref.~\cite{Werner2010} or Fig.~2 in Ref.~\cite{Eckel2010}.
The disordered cases yield a reasonable interpolation between these two limits.
Remarkably, even though the DOS peak in the double quantum dot system is not situated at $\omega=0$ and there is no pinning of a Kondo resonance, we observe that the effect of the Coulomb interaction on the current is very small for small $V$, while it becomes pronounced for large voltage bias.
\begin{figure}
\includegraphics[width=1.0\linewidth]{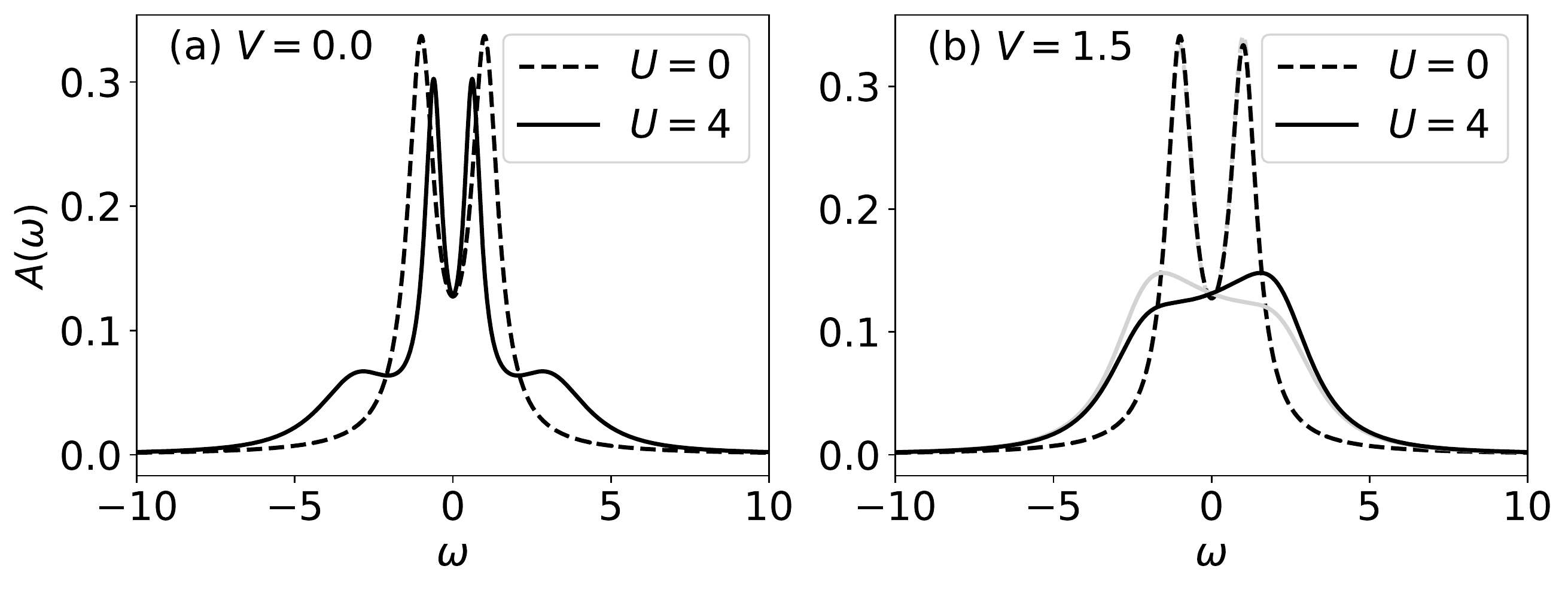}
\caption{Spectral function of the host atoms ($U=0$, dashed line) and impurity atoms ($U=4$, solid line) at (a) equilibrium and (b) in the presence of a bias voltage $V=1.5$.
\label{fig: chain_spc}}
\end{figure}

To gain more insights into the IV characteristics, we plot in Fig.~\ref{fig: chain_spc} the spectral functions for (a) $V=0$ and (b) $V=1.5$. The dashed and solid lines correspond to $U=0$ and $U=4$, respectively. 
Note that the central scattering region contains two sites, and the local spectral function of the left and right site are symmetric with respect to $\omega\rightarrow -\omega$ due to electron-hole symmetry (black solid line for the left site, light grey for the right site).
In the equilibrium case with $U=0$, displayed in Fig.~\ref{fig: chain_spc}(a), the DOS shows two peaks (dashed line), representing the bonding and anti-bonding states formed by the hopping between the two central sites.
As the interaction strength increases to $U=4$, the peaks shift slightly towards each other, and their intensity decreases as some weight is transferred to shoulder structures located around $\omega = \pm 3$. These structures are further analyzed in Appendix \ref{sec: spectra of hubbard dimer}, which presents exact diagonalization results that qualitatively reproduce both spectra. 
When a voltage bias of $V=1.5$ is applied, as shown in Fig.~\ref{fig: chain_spc}~(b), the non-interacting spectral function (dashed lines) remains largely unaffected by the bias.
However, when $U=4$ (panel (b)), the peaks associated with the bonding and anti-bonding states are quickly suppressed and eventually merge with the shoulder features in the presence of the bias.
It is noteworthy that the spectrum's value at zero frequency remains largely unaffected by the el-el interactions. 
Due to the Meir-Wingreen formula \cite{Meir1992,Haug2008}, this characteristic renders the current insensitive to el-el interactions at low bias values,  as shown in Fig.~\ref{fig: chain_IV}.

The properties of the spectral functions explain further aspects of the IV characteristics in Fig.~\ref{fig: chain_IV}. 
In particular, for $U=0$, the current grows faster than linear at small voltages since more bonding and anti-bonding states become available for transport as the bias increases up to approximately $V\approx 1$. 
After that point, the non-interacting current begins to saturate due to the finite width of the density of states.
For $U=4$, the current is expected to saturate only at $V\approx 5$ due to the wider density of states, as shown in Fig.~\ref{fig: chain_spc}~(b).
However, as the bias is increased to $V\approx 1.5$, more electronic states become involved in the transport mechanism (Fig.~\ref{fig: chain_spc}~(b)), which also leads to an upturn in the current-voltage characteristic of the interacting system.  

\section{Conclusions\label{sec: conclusions}}

We presented a nonequilibrium DMFT approach to inhomogeneous Hubbard-Anderson lattice models that treats disorders and electron-electron interactions on equal footing.
The theory reduces to the conventional DMFT for ordered lattices and to the CPA for non-interacting electrons. 
Both time-dependent and steady-state problems have been implemented with advanced numerical methods.

To validate our approach, we performed benchmarks on an isolated cube composed of eight sites. These test calculations demonstrated a good agreement of the spectral functions with exact diagonalization results at both weak and strong couplings, and for a wide range of disorder concentrations.
Furthermore, we investigated an interaction quench problem and showed that our scheme almost perfectly conserves the total energy during the time evolution, and that both the clean and disordered systems thermalize within just a few hopping times.

As an application, we studied a quantum transport model featuring a serial double quantum dot between two leads. This model includes both disorders and electron-electron interactions within the device region.
We found that the occupations on the left and right dots can be reversed as a function of external parameters, such as voltage or interaction strength.
Moreover, we showed that interactions suppress the current at large bias values, while their effect on the current becomes negligible at low bias.
Additionally, we discussed how the electron-electron interactions redistribute the electron population, leading to an effectively hot non-equilibrium steady-state.
Models with disorder yield a reasonable interpolation between the results for the clean host and impurity systems.
Our method offers a versatile framework for studying nonequilibrium phenomena in which both interaction and disorder effects play an important role.

Although our study primarily focused on systems close to particle-hole symmetry, due to the limitations of the employed IPT impurity solver, it is straightforward to incorporate more advanced impurity solvers to explore a broader parameter regime. In equilibrium, Monte Carlo solvers are a natural choice \cite{Rubtsov2005,Werner2006}, while nonequilibrium simulations of strongly correlated systems could be implemented with perturbative strong-coupling solvers \cite{Eckstein2010b}. Furthermore, our theory can be combined with a realistic orbital basis set to realize first-principles simulations of transport or other nonequilibrium properties.

\begin{acknowledgments}
The calculations have been run on the Beo05 cluster at the University of Fribourg. We acknowledge support from ERC Consolidator Grant No.~724103 and SNSF Grant No. 200021-196966.
\end{acknowledgments}

\appendix

\section{Lead self-energy and gauge transformation}\label{sec: lead self-energy}

In quantum transport problems, the system is coupled to external fermionic baths, whose effects can be incorporated into the lead self-energy, as was done in Eq.~\eqref{eq: lattice action}.
We assume that the leads are non-interacting and free of disorder.
The Hamiltonian of the total system takes the form
\begin{subequations}\label{eq: transport Hamiltonian}
\begin{dmath}
\hat{H}(t) = \sum_{\alpha} \left( \hat{H}^{\text{$\alpha$-ld}}(t) + \hat{H}^{\text{$\alpha$-hyb}}(t) \right) + \hat{H}^{\text{dev}}(t)~,
\end{dmath}
where 
\begin{dmath}
\hat{H}^{\text{$\alpha$-ld}}(t) = \sum_{mn,\sigma} H^{\text{$\alpha$-ld}}_{mn,\sigma}(t) a_{\alpha m\sigma}^\dag a_{\alpha n\sigma}~,
\end{dmath}
\begin{dmath}
\hat{H}^{\text{$\alpha$-hyb}}(t) = \sum_{im,\sigma} \left( H^{\text{$\alpha$-hyb}}_{im,\sigma}(t) c_{i\sigma}^\dag a_{\alpha m\sigma} + h.c.
 \right)~.
\end{dmath}
\end{subequations}
Here, $\hat{H}^{\text{$\alpha$-ld}}(t)$ and $\hat{H}^{\text{$\alpha$-hyb}}(t)$ are the Hamiltonian for the $\alpha$-lead and its coupling to the central device region, respectively.
$\hat{H}^{\text{dev}}(t)$ is the device Hamiltonian, whose explicit form is not relevant for the lead self-energy calculation.
We use $a$ ($a^\dag$) to denote the annihilation (creation) operators of the lead electrons.
Due to the hermiticity of the Hamiltonian, $H_{mn,\sigma}^{\text{$\alpha$-ld}}(t) = [H_{nm,\sigma}^{\text{$\alpha$-ld}}(t)]^* $.

\subsection{Lead self-energy}

The action associated with the Hamiltonian \eqref{eq: transport Hamiltonian} reads
\begin{dmath}\label{eq: transport Hamiltonian action}
S = \int dz \left\{ \sum_{\alpha,mn,\sigma} a^*_{\alpha m\sigma}(z) \left( i\frac{\overrightarrow{d}}{dz} - H_{mn,\sigma}^{\text{$\alpha$-ld}}(z) \right) a_{\alpha n\sigma}(z) - \sum_{\alpha,im,\sigma} \left( c_{i\sigma}^*(z) H_{im,\sigma}^{\text{$\alpha$-hyb}}(z) a_{\alpha m\sigma}(z) + h.c. \right) \right\} + S^{\text{dev}}~.
\end{dmath}
Since the leads are noninteracting, one can integrate them out using a Gaussian integral, which results in an effective action $S^{\text{eff}}$ for the central device region.
If an observable $A[c^*,c]$ is defined on the device subspace, its expectation value reads
\begin{dmath}\label{eq: effective action}
A = \frac{\int \mathcal{D}[a_\alpha^*,a_\alpha; c^*, c] A[c^*,c] e^{iS[a_\alpha^*,a_\alpha; c^*, c]}}{\int \mathcal{D}[a_\alpha^*,a_\alpha; c^*, c] e^{iS[a_\alpha^*,a_\alpha; c^*, c]}}
= \frac{\int \mathcal{D}[c^*, c] A[c^*,c] e^{iS^{\text{eff}}[c^*, c]}}{\int \mathcal{D}[c^*, c] e^{iS^{\text{eff}}[c^*, c]}}~,
\end{dmath}
where the second line defines $S^{\text{eff}}[c^*,c]$. 

By inserting Eq.~\eqref{eq: transport Hamiltonian action} into Eq.~\eqref{eq: effective action} and integrating over $a_{\alpha}$ and $a_{\alpha}^*$, we arrive at
\begin{dmath}\label{eq: effective action for device}
S^{\text{eff}} = S^{\text{dev}} - \int dz dz' \sum_{ij,\sigma} c^*_{i \sigma}(z) 
\sum_\alpha \Sigma_{ij,\sigma}^{\text{$\alpha$-ld}}(z,z')
c_{j \sigma}(z')~,
\end{dmath}
where the (embedding) self-energy of the $\alpha$-lead reads
\begin{dmath}\label{eq: lead self-energy}
\Sigma^{\text{$\alpha$-ld}}_{ij,\sigma}(z,z') = \sum_{kl } \left[ H_{ik,\sigma}^{\text{$\alpha$-hyb}}(z) G_{kl,\sigma}^{0,\text{$\alpha$-ld}}(z,z') H_{lj,\sigma}^{\text{$\alpha$-hyb} \,*}(z') \right]~.
\end{dmath}
$G^0$ in the above expression is the Green's function of the decoupled lead $\alpha$:
\begin{dmath}\label{eq: decoupled lead Green's function}
\left( i\frac{\overrightarrow{d}}{dz} - H_{ml,\sigma}^{\text{$\alpha$-ld}}(z) \right) G^{0,\text{$\alpha$-ld}}_{ln,\sigma}(z,z') = \delta(z-z')\delta_{mn}~.
\end{dmath}
Comparing Eq.~\eqref{eq: effective action for device} with Eq.~\eqref{eq: lattice action}, we see that the external self-energy from the leads is given by $\Sigma^{\text{ext}} = \sum_\alpha \Sigma^{\text{$\alpha$-ld}}$.

\subsection{Gauge transformation}

A time-dependent external bias $V^{\alpha}(t)$ shifts the on-site energy of the $\alpha$-lead, i.e. the matrix elements in the occupation number basis are shifted as
$H^{\text{$\alpha$-ld}}_{mn,\sigma}(t) = H^{\text{$\alpha$-ld}}_{mn,\sigma}(t_0) + \theta(t - t_0) V^{\alpha}(t) \delta_{mn}$. 
One can employ a gauge transformation to shift the time-dependence from the onsite energy of the lead to the hybridization term, which can simplify the calculation of the lead self-energy since the Green's function of the isolated lead is in this case time-independent.

In practice, this is achieved by applying a time-dependent transformation
\begin{dmath}
\hat{\tilde{H}}(t) = U(t) \left( \hat{H}(t) - i\frac{\partial}{\partial t} \right) U^\dag(t)
\end{dmath}
to the Hamiltonian \eqref{eq: transport Hamiltonian} \cite{PhysRevB.74.085324}, 
with the unitary operator $U(t)$ given by
\begin{dmath}
\hat{U}(t) = \exp{\left( i\int_{t_0}^t d\bar{t} \sum_{\alpha m \sigma} V^{\alpha}(\bar{t}) \hat{n}_{\alpha m\sigma} \right)}~.
\end{dmath}
The transformed wave function becomes $\ket{\tilde{\Psi}(t)} = \hat{U}(t) \ket{\Psi(t)}$, and its time evolution is determined by the Schr\"odinger equation
\begin{dmath}
i\frac{\partial}{\partial t}\ket{\tilde{\Psi}(t)} = \hat{\tilde{H}}(t) \ket{\tilde{\Psi}(t)}~.
\end{dmath}
One can prove that physical observables $A(t) = \avg{\tilde{\Psi}(t)|\hat{A}(t)|\tilde{\Psi}(t)}$ evaluated with the new wave function $\ket{\tilde{\Psi}(t)}$ are the same as in the original formulation. 

After the transformation, the Hamiltonian \eqref{eq: transport Hamiltonian} becomes
\begin{subequations}\label{eq: transport Hamiltonian after gauge transform}
\begin{dmath}
\hat{\tilde{H}}(t) = \sum_{\alpha} \left( \hat{\tilde{H}}^{\text{$\alpha$-ld}} + \hat{\tilde{H}}^{\text{$\alpha$-hyb}}(t) \right) + \hat{\tilde{H}}^{\text{dev}}(t)~,
\end{dmath}
where
\begin{dmath}
\hat{\tilde{H}}^{\text{$\alpha$-ld}} = \sum_{mn,\sigma} H^{\text{$\alpha$-ld}}_{mn,\sigma}(t_0) a_{\alpha m\sigma}^\dag a_{\alpha n\sigma}~,
\end{dmath}
\begin{dmath}
\hat{\tilde{H}}^{\text{$\alpha$-hyb}}(t) = \sum_{im,\sigma} \left( H^{\text{$\alpha$-hyb}}_{im,\sigma}(t) e^{-i\int_{t_0}^t d\bar{t}V^\alpha(\bar{t})} c_{i\sigma}^\dag a_{\alpha m\sigma} + h.c.
 \right),
\end{dmath}
\end{subequations}
and $\hat{\tilde{H}}^{\text{dev}}(t) = \hat{H}^{\text{dev}}(t)$.
In deriving Eq.~\eqref{eq: transport Hamiltonian after gauge transform}, one uses $\hat{U}(t) a_{\alpha m\sigma} \hat{U}^\dag(t) = a_{\alpha m\sigma} e^{- i\int_{t_0}^t d\bar{t} V^\alpha(\bar{t})}$ and  $\hat{U}(t) a^\dag_{\alpha m\sigma} \hat{U}^\dag(t) = a^\dag_{\alpha m\sigma} e^{+ i\int_{t_0}^t d\bar{t} V^\alpha(\bar{t})}$, which can be obtained from the Baker–Campbell–Hausdorff formula.
Note that in the transformed Hamiltonian, Eq.~\eqref{eq: transport Hamiltonian after gauge transform}, the lead Hamiltonian is time-independent, which enables us to apply conventional (time-independent) recursive methods to calculate the decoupled lead Green's function in Eq.~\eqref{eq: decoupled lead Green's function} \cite{Godfrin1991,*Sancho1985}.

\section{Lattice hybridization function}\label{sec: lattice hybridization function}

In this appendix, we provide a proof for Eq.~\eqref{eq: lattice hybridization function} based on the idea presented in Ref.~\cite{gonis1992green}.
To simplify our notation, we exclude unnecessary indices, and keep only the site index denoted by $i,l,\ldots$ in this section.
Repeated indices are summed over. 
From Eq.~\eqref{eq: lattice Dyson equation}, the site diagonal part of $\Gamma$ reads ($\tilde{W}_{ii}$ generically is non-zero due to the external self-energy)
\begin{dmath}\label{eq: diagonal Gamma}
\Gamma_{ii} = \gamma_i +\gamma_i \tilde{W}_{ii}\Gamma_{ii} +  \gamma_i \tilde{W}_{il}\Gamma_{li}( 1- \delta_{il}).
\end{dmath}
The off-diagonal part of $\Gamma$ is ($l\neq i$)
\begin{dmath}\label{eq: iterative series for Gamma}
\Gamma_{li} 
= \gamma_l \tilde{W}_{lm} \Gamma_{mi}
= \gamma_l \tilde{W}_{li} \Gamma_{ii} + \gamma_l \tilde{W}_{lm} \Gamma_{mi} (1 - \delta_{im})
= \gamma_l \tilde{W}_{li} \Gamma_{ii} + \gamma_l \tilde{W}_{lm} \gamma_{m} \tilde{W}_{mi} \Gamma_{ii} (1-\delta_{im}) + \gamma_l \tilde{W}_{lm} \gamma_m \tilde{W}_{mn} \Gamma_{ni} (1-\delta_{in}) (1-\delta_{im})
= \gamma_l \tilde{W}_{li} \Gamma_{ii} + \gamma_l \tilde{W}_{lm} \gamma_{m} \tilde{W}_{mi} \Gamma_{ii} (1-\delta_{im}) + \gamma_l \tilde{W}_{lm} \gamma_m \tilde{W}_{mn} \gamma_n \tilde{W}_{ni} \Gamma_{ii} (1-\delta_{in}) (1-\delta_{im})
+ \cdots
\end{dmath}
By inserting Eq.~\eqref{eq: iterative series for Gamma} into Eq.~\eqref{eq: diagonal Gamma} and comparing the result with Eq.~\eqref{eq: impurity dyson equation}, one obtains Eq.~\eqref{eq: lattice hybridization function}.\\[0.5cm]

\begin{figure}[b]
\includegraphics[width=1.0\linewidth]{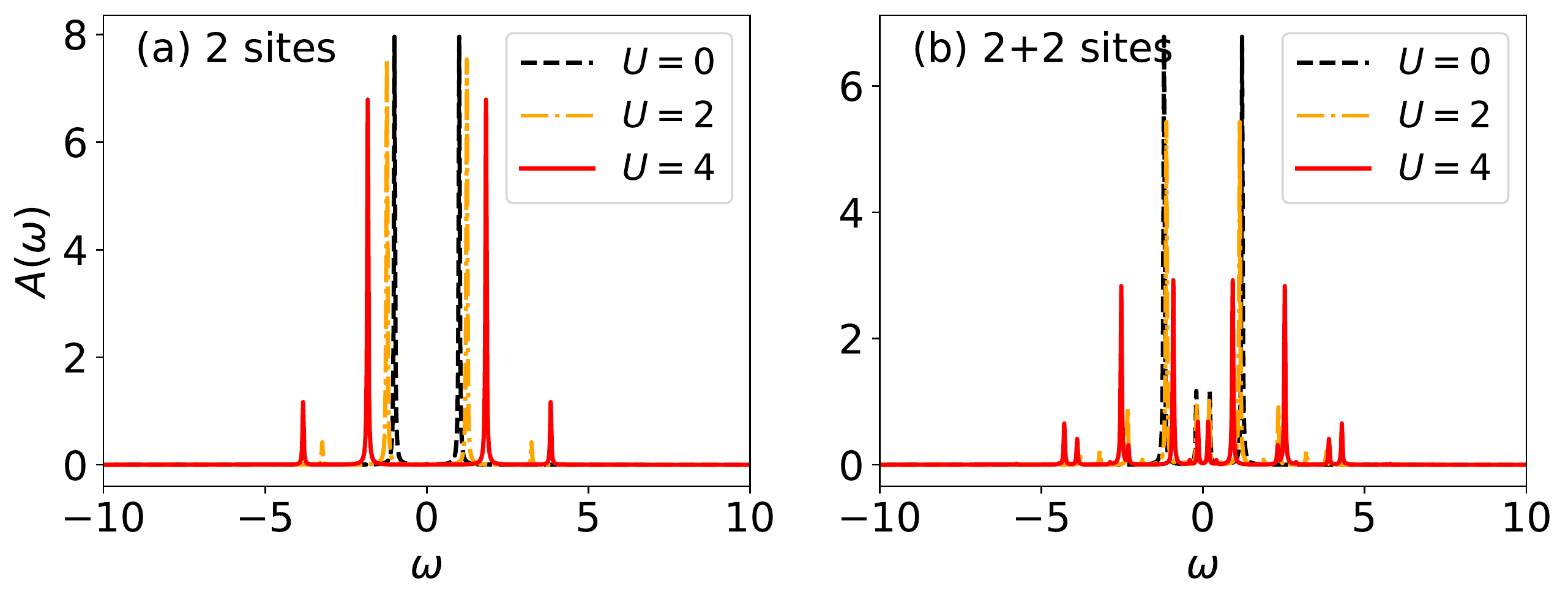}
\caption{Exact local spectral function of (a) a half-filled isolated Hubbard dimer with hopping 1 and (b) the same Hubbard dimer coupled to two non-interacting sites (one on each side, hopping 0.5).
\label{fig: chain_spc_ed}}
\end{figure}

\section{Exact spectra of a Hubbard dimer}\label{sec: spectra of hubbard dimer}

In Fig. \ref{fig: chain_spc_ed}~(a) and (b), we plot the local spectral functions on either the left or right site (the results are identical) for an isolated Hubbard dimer and a dimer coupled to non-interacting sites (one on each side). 
Half-filling is assumed and the intra-dimer hopping amplitude is set to unity. In panel (b), the hopping to the non-interacting sites is $0.5$.
The black, orange, and red lines correspond to $U=0$, $U=2$, and $U=4$, respectively.

The non-interacting spectra shown in panel (a) display two peaks corresponding to bonding and anti-bonding states. Increasing $U$ moves these peaks further apart, with two additional peaks appearing at higher frequencies. This behavior is not consistent with the evolution of the spectra with interaction strength shown in Fig.~\ref{fig: chain_spc}~(a). To understand this behavior, one needs to consider the effect of the coupling to the leads. 
The spectra for the dimer coupled to two noninteracting sites, shown in panel (b), exhibits four peaks at $U=0$: two main peaks near the bonding and anti-bonding states of panel (a), and two smaller peaks near the Fermi energy, which fill in the gap. Increasing $U$ causes the main peaks to approach each other, while spectral weight is transferred more rapidly to higher energy states, compared to the isolated case.
This behavior qualitatively explains the evolution of the DMFT equilibrium spectra with interaction strength, shown in Fig.~\ref{fig: chain_spc}~(a).

\bibliography{ref}

\begin{thebibliography}{65}%
\makeatletter
\providecommand \@ifxundefined [1]{%
 \@ifx{#1\undefined}
}%
\providecommand \@ifnum [1]{%
 \ifnum #1\expandafter \@firstoftwo
 \else \expandafter \@secondoftwo
 \fi
}%
\providecommand \@ifx [1]{%
 \ifx #1\expandafter \@firstoftwo
 \else \expandafter \@secondoftwo
 \fi
}%
\providecommand \natexlab [1]{#1}%
\providecommand \enquote  [1]{``#1''}%
\providecommand \bibnamefont  [1]{#1}%
\providecommand \bibfnamefont [1]{#1}%
\providecommand \citenamefont [1]{#1}%
\providecommand \href@noop [0]{\@secondoftwo}%
\providecommand \href [0]{\begingroup \@sanitize@url \@href}%
\providecommand \@href[1]{\@@startlink{#1}\@@href}%
\providecommand \@@href[1]{\endgroup#1\@@endlink}%
\providecommand \@sanitize@url [0]{\catcode `\\12\catcode `\$12\catcode
  `\&12\catcode `\#12\catcode `\^12\catcode `\_12\catcode `\%12\relax}%
\providecommand \@@startlink[1]{}%
\providecommand \@@endlink[0]{}%
\providecommand \url  [0]{\begingroup\@sanitize@url \@url }%
\providecommand \@url [1]{\endgroup\@href {#1}{\urlprefix }}%
\providecommand \urlprefix  [0]{URL }%
\providecommand \Eprint [0]{\href }%
\providecommand \doibase [0]{https://doi.org/}%
\providecommand \selectlanguage [0]{\@gobble}%
\providecommand \bibinfo  [0]{\@secondoftwo}%
\providecommand \bibfield  [0]{\@secondoftwo}%
\providecommand \translation [1]{[#1]}%
\providecommand \BibitemOpen [0]{}%
\providecommand \bibitemStop [0]{}%
\providecommand \bibitemNoStop [0]{.\EOS\space}%
\providecommand \EOS [0]{\spacefactor3000\relax}%
\providecommand \BibitemShut  [1]{\csname bibitem#1\endcsname}%
\let\auto@bib@innerbib\@empty
\bibitem [{\citenamefont {Belitz}\ and\ \citenamefont
  {Kirkpatrick}(1994)}]{Belitz1994}%
  \BibitemOpen
  \bibfield  {author} {\bibinfo {author} {\bibfnamefont {D.}~\bibnamefont
  {Belitz}}\ and\ \bibinfo {author} {\bibfnamefont {T.~R.}\ \bibnamefont
  {Kirkpatrick}},\ }\bibfield  {title} {\bibinfo {title} {The anderson-mott
  transition},\ }\href {https://doi.org/10.1103/RevModPhys.66.261} {\bibfield
  {journal} {\bibinfo  {journal} {Rev. Mod. Phys.}\ }\textbf {\bibinfo {volume}
  {66}},\ \bibinfo {pages} {261} (\bibinfo {year} {1994})}\BibitemShut
  {NoStop}%
\bibitem [{\citenamefont {Lee}\ and\ \citenamefont
  {Ramakrishnan}(1985)}]{Lee1985}%
  \BibitemOpen
  \bibfield  {author} {\bibinfo {author} {\bibfnamefont {P.~A.}\ \bibnamefont
  {Lee}}\ and\ \bibinfo {author} {\bibfnamefont {T.~V.}\ \bibnamefont
  {Ramakrishnan}},\ }\bibfield  {title} {\bibinfo {title} {Disordered
  electronic systems},\ }\href {https://doi.org/10.1103/RevModPhys.57.287}
  {\bibfield  {journal} {\bibinfo  {journal} {Rev. Mod. Phys.}\ }\textbf
  {\bibinfo {volume} {57}},\ \bibinfo {pages} {287} (\bibinfo {year}
  {1985})}\BibitemShut {NoStop}%
\bibitem [{\citenamefont {MOTT}(1968)}]{N.F.1968}%
  \BibitemOpen
  \bibfield  {author} {\bibinfo {author} {\bibfnamefont {N.~F.}\ \bibnamefont
  {MOTT}},\ }\bibfield  {title} {\bibinfo {title} {Metal-insulator
  transition},\ }\href {https://doi.org/10.1103/RevModPhys.40.677} {\bibfield
  {journal} {\bibinfo  {journal} {Rev. Mod. Phys.}\ }\textbf {\bibinfo {volume}
  {40}},\ \bibinfo {pages} {677} (\bibinfo {year} {1968})}\BibitemShut
  {NoStop}%
\bibitem [{\citenamefont {Lee}\ \emph {et~al.}(2006)\citenamefont {Lee},
  \citenamefont {Nagaosa},\ and\ \citenamefont {Wen}}]{Lee2006}%
  \BibitemOpen
  \bibfield  {author} {\bibinfo {author} {\bibfnamefont {P.~A.}\ \bibnamefont
  {Lee}}, \bibinfo {author} {\bibfnamefont {N.}~\bibnamefont {Nagaosa}},\ and\
  \bibinfo {author} {\bibfnamefont {X.-G.}\ \bibnamefont {Wen}},\ }\bibfield
  {title} {\bibinfo {title} {Doping a mott insulator: Physics of
  high-temperature superconductivity},\ }\href
  {https://doi.org/10.1103/RevModPhys.78.17} {\bibfield  {journal} {\bibinfo
  {journal} {Rev. Mod. Phys.}\ }\textbf {\bibinfo {volume} {78}},\ \bibinfo
  {pages} {17} (\bibinfo {year} {2006})}\BibitemShut {NoStop}%
\bibitem [{\citenamefont {Parkin}\ \emph {et~al.}(2004)\citenamefont {Parkin},
  \citenamefont {Kaiser}, \citenamefont {Panchula}, \citenamefont {Rice},
  \citenamefont {Hughes}, \citenamefont {Samant},\ and\ \citenamefont
  {Yang}}]{Parkin2004}%
  \BibitemOpen
  \bibfield  {author} {\bibinfo {author} {\bibfnamefont {S.~S.~P.}\
  \bibnamefont {Parkin}}, \bibinfo {author} {\bibfnamefont {C.}~\bibnamefont
  {Kaiser}}, \bibinfo {author} {\bibfnamefont {A.}~\bibnamefont {Panchula}},
  \bibinfo {author} {\bibfnamefont {P.~M.}\ \bibnamefont {Rice}}, \bibinfo
  {author} {\bibfnamefont {B.}~\bibnamefont {Hughes}}, \bibinfo {author}
  {\bibfnamefont {M.}~\bibnamefont {Samant}},\ and\ \bibinfo {author}
  {\bibfnamefont {S.-H.}\ \bibnamefont {Yang}},\ }\bibfield  {title} {\bibinfo
  {title} {Giant tunnelling magnetoresistance at room temperature with mgo
  (100) tunnel barriers},\ }\href {https://doi.org/10.1038/nmat1256} {\bibfield
   {journal} {\bibinfo  {journal} {Nature Materials}\ }\textbf {\bibinfo
  {volume} {3}},\ \bibinfo {pages} {862} (\bibinfo {year} {2004})}\BibitemShut
  {NoStop}%
\bibitem [{\citenamefont {Abanin}\ \emph {et~al.}(2019)\citenamefont {Abanin},
  \citenamefont {Altman}, \citenamefont {Bloch},\ and\ \citenamefont
  {Serbyn}}]{Abanin2019}%
  \BibitemOpen
  \bibfield  {author} {\bibinfo {author} {\bibfnamefont {D.~A.}\ \bibnamefont
  {Abanin}}, \bibinfo {author} {\bibfnamefont {E.}~\bibnamefont {Altman}},
  \bibinfo {author} {\bibfnamefont {I.}~\bibnamefont {Bloch}},\ and\ \bibinfo
  {author} {\bibfnamefont {M.}~\bibnamefont {Serbyn}},\ }\bibfield  {title}
  {\bibinfo {title} {Colloquium: Many-body localization, thermalization, and
  entanglement},\ }\href {https://doi.org/10.1103/RevModPhys.91.021001}
  {\bibfield  {journal} {\bibinfo  {journal} {Rev. Mod. Phys.}\ }\textbf
  {\bibinfo {volume} {91}},\ \bibinfo {pages} {021001} (\bibinfo {year}
  {2019})}\BibitemShut {NoStop}%
\bibitem [{\citenamefont {Zwanenburg}\ \emph {et~al.}(2013)\citenamefont
  {Zwanenburg}, \citenamefont {Dzurak}, \citenamefont {Morello}, \citenamefont
  {Simmons}, \citenamefont {Hollenberg}, \citenamefont {Klimeck}, \citenamefont
  {Rogge}, \citenamefont {Coppersmith},\ and\ \citenamefont
  {Eriksson}}]{Zwanenburg2013}%
  \BibitemOpen
  \bibfield  {author} {\bibinfo {author} {\bibfnamefont {F.~A.}\ \bibnamefont
  {Zwanenburg}}, \bibinfo {author} {\bibfnamefont {A.~S.}\ \bibnamefont
  {Dzurak}}, \bibinfo {author} {\bibfnamefont {A.}~\bibnamefont {Morello}},
  \bibinfo {author} {\bibfnamefont {M.~Y.}\ \bibnamefont {Simmons}}, \bibinfo
  {author} {\bibfnamefont {L.~C.~L.}\ \bibnamefont {Hollenberg}}, \bibinfo
  {author} {\bibfnamefont {G.}~\bibnamefont {Klimeck}}, \bibinfo {author}
  {\bibfnamefont {S.}~\bibnamefont {Rogge}}, \bibinfo {author} {\bibfnamefont
  {S.~N.}\ \bibnamefont {Coppersmith}},\ and\ \bibinfo {author} {\bibfnamefont
  {M.~A.}\ \bibnamefont {Eriksson}},\ }\bibfield  {title} {\bibinfo {title}
  {Silicon quantum electronics},\ }\href
  {https://doi.org/10.1103/RevModPhys.85.961} {\bibfield  {journal} {\bibinfo
  {journal} {Rev. Mod. Phys.}\ }\textbf {\bibinfo {volume} {85}},\ \bibinfo
  {pages} {961} (\bibinfo {year} {2013})}\BibitemShut {NoStop}%
\bibitem [{\citenamefont {Bruus}\ and\ \citenamefont
  {Flensberg}(2004)}]{Bruus2004}%
  \BibitemOpen
  \bibfield  {author} {\bibinfo {author} {\bibfnamefont {H.}~\bibnamefont
  {Bruus}}\ and\ \bibinfo {author} {\bibfnamefont {K.}~\bibnamefont
  {Flensberg}},\ }\href@noop {} {\emph {\bibinfo {title} {Many-body quantum
  theory in condensed matter physics: an introduction}}}\ (\bibinfo
  {publisher} {Oxford university press},\ \bibinfo {year} {2004})\BibitemShut
  {NoStop}%
\bibitem [{\citenamefont {Mahan}(2000)}]{Mahan2000}%
  \BibitemOpen
  \bibfield  {author} {\bibinfo {author} {\bibfnamefont {G.~D.}\ \bibnamefont
  {Mahan}},\ }\href@noop {} {\emph {\bibinfo {title} {Many-particle physics}}}\
  (\bibinfo  {publisher} {Springer Science \& Business Media},\ \bibinfo {year}
  {2000})\BibitemShut {NoStop}%
\bibitem [{\citenamefont {Okhotnikov}\ \emph {et~al.}(2016)\citenamefont
  {Okhotnikov}, \citenamefont {Charpentier},\ and\ \citenamefont
  {Cadars}}]{Okhotnikov2016}%
  \BibitemOpen
  \bibfield  {author} {\bibinfo {author} {\bibfnamefont {K.}~\bibnamefont
  {Okhotnikov}}, \bibinfo {author} {\bibfnamefont {T.}~\bibnamefont
  {Charpentier}},\ and\ \bibinfo {author} {\bibfnamefont {S.}~\bibnamefont
  {Cadars}},\ }\bibfield  {title} {\bibinfo {title} {Supercell program: a
  combinatorial structure-generation approach for the local-level modeling of
  atomic substitutions and partial occupancies in crystals},\ }\href
  {https://doi.org/10.1186/s13321-016-0129-3} {\bibfield  {journal} {\bibinfo
  {journal} {Journal of Cheminformatics}\ }\textbf {\bibinfo {volume} {8}},\
  \bibinfo {pages} {17} (\bibinfo {year} {2016})}\BibitemShut {NoStop}%
\bibitem [{\citenamefont {Datta}(2005)}]{Datta2005}%
  \BibitemOpen
  \bibfield  {author} {\bibinfo {author} {\bibfnamefont {S.}~\bibnamefont
  {Datta}},\ }\href@noop {} {\emph {\bibinfo {title} {Quantum transport: atom
  to transistor}}}\ (\bibinfo  {publisher} {Cambridge University Press},\
  \bibinfo {year} {2005})\BibitemShut {NoStop}%
\bibitem [{\citenamefont {Datta}(1997)}]{Datta1997}%
  \BibitemOpen
  \bibfield  {author} {\bibinfo {author} {\bibfnamefont {S.}~\bibnamefont
  {Datta}},\ }\href@noop {} {\emph {\bibinfo {title} {Electronic transport in
  mesoscopic systems}}}\ (\bibinfo  {publisher} {Cambridge university press},\
  \bibinfo {year} {1997})\BibitemShut {NoStop}%
\bibitem [{\citenamefont {Gubernatis}\ \emph {et~al.}(2016)\citenamefont
  {Gubernatis}, \citenamefont {Kawashima},\ and\ \citenamefont
  {Werner}}]{Gubernatis2016}%
  \BibitemOpen
  \bibfield  {author} {\bibinfo {author} {\bibfnamefont {J.}~\bibnamefont
  {Gubernatis}}, \bibinfo {author} {\bibfnamefont {N.}~\bibnamefont
  {Kawashima}},\ and\ \bibinfo {author} {\bibfnamefont {P.}~\bibnamefont
  {Werner}},\ }\href@noop {} {\emph {\bibinfo {title} {Quantum Monte Carlo
  Methods}}}\ (\bibinfo  {publisher} {Cambridge University Press},\ \bibinfo
  {year} {2016})\BibitemShut {NoStop}%
\bibitem [{\citenamefont {Schollw\"ock}(2005)}]{Schollwoeck2005}%
  \BibitemOpen
  \bibfield  {author} {\bibinfo {author} {\bibfnamefont {U.}~\bibnamefont
  {Schollw\"ock}},\ }\bibfield  {title} {\bibinfo {title} {The density-matrix
  renormalization group},\ }\href {https://doi.org/10.1103/RevModPhys.77.259}
  {\bibfield  {journal} {\bibinfo  {journal} {Rev. Mod. Phys.}\ }\textbf
  {\bibinfo {volume} {77}},\ \bibinfo {pages} {259} (\bibinfo {year}
  {2005})}\BibitemShut {NoStop}%
\bibitem [{\citenamefont {Caffarel}\ and\ \citenamefont
  {Krauth}(1994)}]{Caffarel1994}%
  \BibitemOpen
  \bibfield  {author} {\bibinfo {author} {\bibfnamefont {M.}~\bibnamefont
  {Caffarel}}\ and\ \bibinfo {author} {\bibfnamefont {W.}~\bibnamefont
  {Krauth}},\ }\bibfield  {title} {\bibinfo {title} {Exact diagonalization
  approach to correlated fermions in infinite dimensions: Mott transition and
  superconductivity},\ }\href {https://doi.org/10.1103/PhysRevLett.72.1545}
  {\bibfield  {journal} {\bibinfo  {journal} {Phys. Rev. Lett.}\ }\textbf
  {\bibinfo {volume} {72}},\ \bibinfo {pages} {1545} (\bibinfo {year}
  {1994})}\BibitemShut {NoStop}%
\bibitem [{\citenamefont {Gonis}(1992)}]{gonis1992green}%
  \BibitemOpen
  \bibfield  {author} {\bibinfo {author} {\bibfnamefont {A.}~\bibnamefont
  {Gonis}},\ }\href@noop {} {\emph {\bibinfo {title} {Green functions for
  ordered and disordered systems}}}\ (\bibinfo {year} {1992})\BibitemShut
  {NoStop}%
\bibitem [{\citenamefont {Zunger}\ \emph {et~al.}(1990)\citenamefont {Zunger},
  \citenamefont {Wei}, \citenamefont {Ferreira},\ and\ \citenamefont
  {Bernard}}]{Zunger1990}%
  \BibitemOpen
  \bibfield  {author} {\bibinfo {author} {\bibfnamefont {A.}~\bibnamefont
  {Zunger}}, \bibinfo {author} {\bibfnamefont {S.-H.}\ \bibnamefont {Wei}},
  \bibinfo {author} {\bibfnamefont {L.~G.}\ \bibnamefont {Ferreira}},\ and\
  \bibinfo {author} {\bibfnamefont {J.~E.}\ \bibnamefont {Bernard}},\
  }\bibfield  {title} {\bibinfo {title} {Special quasirandom structures},\
  }\href {https://doi.org/10.1103/PhysRevLett.65.353} {\bibfield  {journal}
  {\bibinfo  {journal} {Phys. Rev. Lett.}\ }\textbf {\bibinfo {volume} {65}},\
  \bibinfo {pages} {353} (\bibinfo {year} {1990})}\BibitemShut {NoStop}%
\bibitem [{\citenamefont {Georges}\ \emph {et~al.}(1996)\citenamefont
  {Georges}, \citenamefont {Kotliar}, \citenamefont {Krauth},\ and\
  \citenamefont {Rozenberg}}]{Georges1996}%
  \BibitemOpen
  \bibfield  {author} {\bibinfo {author} {\bibfnamefont {A.}~\bibnamefont
  {Georges}}, \bibinfo {author} {\bibfnamefont {G.}~\bibnamefont {Kotliar}},
  \bibinfo {author} {\bibfnamefont {W.}~\bibnamefont {Krauth}},\ and\ \bibinfo
  {author} {\bibfnamefont {M.~J.}\ \bibnamefont {Rozenberg}},\ }\bibfield
  {title} {\bibinfo {title} {Dynamical mean-field theory of strongly correlated
  fermion systems and the limit of infinite dimensions},\ }\href
  {https://doi.org/10.1103/RevModPhys.68.13} {\bibfield  {journal} {\bibinfo
  {journal} {Rev. Mod. Phys.}\ }\textbf {\bibinfo {volume} {68}},\ \bibinfo
  {pages} {13} (\bibinfo {year} {1996})}\BibitemShut {NoStop}%
\bibitem [{\citenamefont {Kotliar}\ \emph {et~al.}(2006)\citenamefont
  {Kotliar}, \citenamefont {Savrasov}, \citenamefont {Haule}, \citenamefont
  {Oudovenko}, \citenamefont {Parcollet},\ and\ \citenamefont
  {Marianetti}}]{Kotliar2006}%
  \BibitemOpen
  \bibfield  {author} {\bibinfo {author} {\bibfnamefont {G.}~\bibnamefont
  {Kotliar}}, \bibinfo {author} {\bibfnamefont {S.~Y.}\ \bibnamefont
  {Savrasov}}, \bibinfo {author} {\bibfnamefont {K.}~\bibnamefont {Haule}},
  \bibinfo {author} {\bibfnamefont {V.~S.}\ \bibnamefont {Oudovenko}}, \bibinfo
  {author} {\bibfnamefont {O.}~\bibnamefont {Parcollet}},\ and\ \bibinfo
  {author} {\bibfnamefont {C.~A.}\ \bibnamefont {Marianetti}},\ }\bibfield
  {title} {\bibinfo {title} {Electronic structure calculations with dynamical
  mean-field theory},\ }\href {https://doi.org/10.1103/RevModPhys.78.865}
  {\bibfield  {journal} {\bibinfo  {journal} {Rev. Mod. Phys.}\ }\textbf
  {\bibinfo {volume} {78}},\ \bibinfo {pages} {865} (\bibinfo {year}
  {2006})}\BibitemShut {NoStop}%
\bibitem [{\citenamefont {Freericks}\ \emph {et~al.}(2006)\citenamefont
  {Freericks}, \citenamefont {Turkowski},\ and\ \citenamefont
  {Zlati\ifmmode~\acute{c}\else \'{c}\fi{}}}]{PhysRevLett.97.266408}%
  \BibitemOpen
  \bibfield  {author} {\bibinfo {author} {\bibfnamefont {J.~K.}\ \bibnamefont
  {Freericks}}, \bibinfo {author} {\bibfnamefont {V.~M.}\ \bibnamefont
  {Turkowski}},\ and\ \bibinfo {author} {\bibfnamefont {V.}~\bibnamefont
  {Zlati\ifmmode~\acute{c}\else \'{c}\fi{}}},\ }\bibfield  {title} {\bibinfo
  {title} {Nonequilibrium dynamical mean-field theory},\ }\href
  {https://doi.org/10.1103/PhysRevLett.97.266408} {\bibfield  {journal}
  {\bibinfo  {journal} {Phys. Rev. Lett.}\ }\textbf {\bibinfo {volume} {97}},\
  \bibinfo {pages} {266408} (\bibinfo {year} {2006})}\BibitemShut {NoStop}%
\bibitem [{\citenamefont {Aoki}\ \emph {et~al.}(2014)\citenamefont {Aoki},
  \citenamefont {Tsuji}, \citenamefont {Eckstein}, \citenamefont {Kollar},
  \citenamefont {Oka},\ and\ \citenamefont {Werner}}]{RevModPhys.86.779}%
  \BibitemOpen
  \bibfield  {author} {\bibinfo {author} {\bibfnamefont {H.}~\bibnamefont
  {Aoki}}, \bibinfo {author} {\bibfnamefont {N.}~\bibnamefont {Tsuji}},
  \bibinfo {author} {\bibfnamefont {M.}~\bibnamefont {Eckstein}}, \bibinfo
  {author} {\bibfnamefont {M.}~\bibnamefont {Kollar}}, \bibinfo {author}
  {\bibfnamefont {T.}~\bibnamefont {Oka}},\ and\ \bibinfo {author}
  {\bibfnamefont {P.}~\bibnamefont {Werner}},\ }\bibfield  {title} {\bibinfo
  {title} {Nonequilibrium dynamical mean-field theory and its applications},\
  }\href {https://doi.org/10.1103/RevModPhys.86.779} {\bibfield  {journal}
  {\bibinfo  {journal} {Rev. Mod. Phys.}\ }\textbf {\bibinfo {volume} {86}},\
  \bibinfo {pages} {779} (\bibinfo {year} {2014})}\BibitemShut {NoStop}%
\bibitem [{\citenamefont {Metzner}\ and\ \citenamefont
  {Vollhardt}(1989)}]{Metzner1989}%
  \BibitemOpen
  \bibfield  {author} {\bibinfo {author} {\bibfnamefont {W.}~\bibnamefont
  {Metzner}}\ and\ \bibinfo {author} {\bibfnamefont {D.}~\bibnamefont
  {Vollhardt}},\ }\bibfield  {title} {\bibinfo {title} {Correlated lattice
  fermions in $d=\ensuremath{\infty}$ dimensions},\ }\href
  {https://doi.org/10.1103/PhysRevLett.62.324} {\bibfield  {journal} {\bibinfo
  {journal} {Phys. Rev. Lett.}\ }\textbf {\bibinfo {volume} {62}},\ \bibinfo
  {pages} {324} (\bibinfo {year} {1989})}\BibitemShut {NoStop}%
\bibitem [{\citenamefont {Soven}(1967)}]{Soven1967}%
  \BibitemOpen
  \bibfield  {author} {\bibinfo {author} {\bibfnamefont {P.}~\bibnamefont
  {Soven}},\ }\bibfield  {title} {\bibinfo {title} {Coherent-potential model of
  substitutional disordered alloys},\ }\href
  {https://doi.org/10.1103/PhysRev.156.809} {\bibfield  {journal} {\bibinfo
  {journal} {Phys. Rev.}\ }\textbf {\bibinfo {volume} {156}},\ \bibinfo {pages}
  {809} (\bibinfo {year} {1967})}\BibitemShut {NoStop}%
\bibitem [{\citenamefont {Velick\'y}\ \emph {et~al.}(1968)\citenamefont
  {Velick\'y}, \citenamefont {Kirkpatrick},\ and\ \citenamefont
  {Ehrenreich}}]{Velicky1968}%
  \BibitemOpen
  \bibfield  {author} {\bibinfo {author} {\bibfnamefont {B.}~\bibnamefont
  {Velick\'y}}, \bibinfo {author} {\bibfnamefont {S.}~\bibnamefont
  {Kirkpatrick}},\ and\ \bibinfo {author} {\bibfnamefont {H.}~\bibnamefont
  {Ehrenreich}},\ }\bibfield  {title} {\bibinfo {title} {Single-site
  approximations in the electronic theory of simple binary alloys},\ }\href
  {https://doi.org/10.1103/PhysRev.175.747} {\bibfield  {journal} {\bibinfo
  {journal} {Phys. Rev.}\ }\textbf {\bibinfo {volume} {175}},\ \bibinfo {pages}
  {747} (\bibinfo {year} {1968})}\BibitemShut {NoStop}%
\bibitem [{\citenamefont {Elliott}\ \emph {et~al.}(1974)\citenamefont
  {Elliott}, \citenamefont {Krumhansl},\ and\ \citenamefont
  {Leath}}]{Elliott1974}%
  \BibitemOpen
  \bibfield  {author} {\bibinfo {author} {\bibfnamefont {R.~J.}\ \bibnamefont
  {Elliott}}, \bibinfo {author} {\bibfnamefont {J.~A.}\ \bibnamefont
  {Krumhansl}},\ and\ \bibinfo {author} {\bibfnamefont {P.~L.}\ \bibnamefont
  {Leath}},\ }\bibfield  {title} {\bibinfo {title} {The theory and properties
  of randomly disordered crystals and related physical systems},\ }\href
  {https://doi.org/10.1103/RevModPhys.46.465} {\bibfield  {journal} {\bibinfo
  {journal} {Rev. Mod. Phys.}\ }\textbf {\bibinfo {volume} {46}},\ \bibinfo
  {pages} {465} (\bibinfo {year} {1974})}\BibitemShut {NoStop}%
\bibitem [{\citenamefont {Jani\ifmmode~\check{s}\else
  \v{s}\fi{}}(1989)}]{Jani1989}%
  \BibitemOpen
  \bibfield  {author} {\bibinfo {author} {\bibfnamefont {V.}~\bibnamefont
  {Jani\ifmmode~\check{s}\else \v{s}\fi{}}},\ }\bibfield  {title} {\bibinfo
  {title} {Free-energy functional in the generalized coherent-potential
  approximation},\ }\href {https://doi.org/10.1103/PhysRevB.40.11331}
  {\bibfield  {journal} {\bibinfo  {journal} {Phys. Rev. B}\ }\textbf {\bibinfo
  {volume} {40}},\ \bibinfo {pages} {11331} (\bibinfo {year}
  {1989})}\BibitemShut {NoStop}%
\bibitem [{\citenamefont {Jani\ifmmode~\check{s}\else \v{s}\fi{}}\ and\
  \citenamefont {Vollhardt}(1992)}]{PhysRevB.46.15712}%
  \BibitemOpen
  \bibfield  {author} {\bibinfo {author} {\bibfnamefont {V.}~\bibnamefont
  {Jani\ifmmode~\check{s}\else \v{s}\fi{}}}\ and\ \bibinfo {author}
  {\bibfnamefont {D.}~\bibnamefont {Vollhardt}},\ }\bibfield  {title} {\bibinfo
  {title} {Coupling of quantum degrees of freedom in strongly interacting
  disordered electron systems},\ }\href
  {https://doi.org/10.1103/PhysRevB.46.15712} {\bibfield  {journal} {\bibinfo
  {journal} {Phys. Rev. B}\ }\textbf {\bibinfo {volume} {46}},\ \bibinfo
  {pages} {15712} (\bibinfo {year} {1992})}\BibitemShut {NoStop}%
\bibitem [{\citenamefont {Janis}\ \emph {et~al.}(1993)\citenamefont {Janis},
  \citenamefont {Ulmke},\ and\ \citenamefont {Vollhardt}}]{Janis1993}%
  \BibitemOpen
  \bibfield  {author} {\bibinfo {author} {\bibfnamefont {V.}~\bibnamefont
  {Janis}}, \bibinfo {author} {\bibfnamefont {M.}~\bibnamefont {Ulmke}},\ and\
  \bibinfo {author} {\bibfnamefont {D.}~\bibnamefont {Vollhardt}},\ }\bibfield
  {title} {\bibinfo {title} {Disorder vs. interaction in the hubbard model:
  Phase diagram in infinite dimensions},\ }\href
  {https://doi.org/10.1209/0295-5075/24/4/009} {\bibfield  {journal} {\bibinfo
  {journal} {Europhysics Letters}\ }\textbf {\bibinfo {volume} {24}},\ \bibinfo
  {pages} {287} (\bibinfo {year} {1993})}\BibitemShut {NoStop}%
\bibitem [{\citenamefont {Ulmke}\ \emph {et~al.}(1995)\citenamefont {Ulmke},
  \citenamefont {Jani\ifmmode~\check{s}\else \v{s}\fi{}},\ and\ \citenamefont
  {Vollhardt}}]{Ulmke1995}%
  \BibitemOpen
  \bibfield  {author} {\bibinfo {author} {\bibfnamefont {M.}~\bibnamefont
  {Ulmke}}, \bibinfo {author} {\bibfnamefont {V.}~\bibnamefont
  {Jani\ifmmode~\check{s}\else \v{s}\fi{}}},\ and\ \bibinfo {author}
  {\bibfnamefont {D.}~\bibnamefont {Vollhardt}},\ }\bibfield  {title} {\bibinfo
  {title} {Anderson-hubbard model in infinite dimensions},\ }\href
  {https://doi.org/10.1103/PhysRevB.51.10411} {\bibfield  {journal} {\bibinfo
  {journal} {Phys. Rev. B}\ }\textbf {\bibinfo {volume} {51}},\ \bibinfo
  {pages} {10411} (\bibinfo {year} {1995})}\BibitemShut {NoStop}%
\bibitem [{\citenamefont {Dobrosavljevi\ifmmode~\acute{c}\else \'{c}\fi{}}\
  and\ \citenamefont {Kotliar}(1997)}]{PhysRevLett.78.3943}%
  \BibitemOpen
  \bibfield  {author} {\bibinfo {author} {\bibfnamefont {V.}~\bibnamefont
  {Dobrosavljevi\ifmmode~\acute{c}\else \'{c}\fi{}}}\ and\ \bibinfo {author}
  {\bibfnamefont {G.}~\bibnamefont {Kotliar}},\ }\bibfield  {title} {\bibinfo
  {title} {Mean field theory of the mott-anderson transition},\ }\href
  {https://doi.org/10.1103/PhysRevLett.78.3943} {\bibfield  {journal} {\bibinfo
   {journal} {Phys. Rev. Lett.}\ }\textbf {\bibinfo {volume} {78}},\ \bibinfo
  {pages} {3943} (\bibinfo {year} {1997})}\BibitemShut {NoStop}%
\bibitem [{\citenamefont {Drchal}\ \emph {et~al.}(1999)\citenamefont {Drchal},
  \citenamefont {Jani\ifmmode~\check{s}\else \v{s}\fi{}},\ and\ \citenamefont
  {Kudrnovsk\'y}}]{Drchal1999}%
  \BibitemOpen
  \bibfield  {author} {\bibinfo {author} {\bibfnamefont {V.}~\bibnamefont
  {Drchal}}, \bibinfo {author} {\bibfnamefont {V.}~\bibnamefont
  {Jani\ifmmode~\check{s}\else \v{s}\fi{}}},\ and\ \bibinfo {author}
  {\bibfnamefont {J.}~\bibnamefont {Kudrnovsk\'y}},\ }\bibfield  {title}
  {\bibinfo {title} {Dynamical electron correlations in weakly interacting
  systems: Tb-lmto approach to metals and random alloys},\ }\href
  {https://doi.org/10.1103/PhysRevB.60.15664} {\bibfield  {journal} {\bibinfo
  {journal} {Phys. Rev. B}\ }\textbf {\bibinfo {volume} {60}},\ \bibinfo
  {pages} {15664} (\bibinfo {year} {1999})}\BibitemShut {NoStop}%
\bibitem [{\citenamefont {Ebert}\ \emph {et~al.}(2011)\citenamefont {Ebert},
  \citenamefont {K{\"o}dderitzsch},\ and\ \citenamefont
  {Min{\'a}r}}]{Ebert2011}%
  \BibitemOpen
  \bibfield  {author} {\bibinfo {author} {\bibfnamefont {H.}~\bibnamefont
  {Ebert}}, \bibinfo {author} {\bibfnamefont {D.}~\bibnamefont
  {K{\"o}dderitzsch}},\ and\ \bibinfo {author} {\bibfnamefont {J.}~\bibnamefont
  {Min{\'a}r}},\ }\bibfield  {title} {\bibinfo {title} {Calculating condensed
  matter properties using the kkr-green's function method---recent developments
  and applications},\ }\href {https://doi.org/10.1088/0034-4885/74/9/096501}
  {\bibfield  {journal} {\bibinfo  {journal} {Reports on Progress in Physics}\
  }\textbf {\bibinfo {volume} {74}},\ \bibinfo {pages} {096501} (\bibinfo
  {year} {2011})}\BibitemShut {NoStop}%
\bibitem [{\citenamefont {Terletska}\ \emph {et~al.}(2013)\citenamefont
  {Terletska}, \citenamefont {Yang}, \citenamefont {Meng}, \citenamefont
  {Moreno},\ and\ \citenamefont {Jarrell}}]{Terletska2013}%
  \BibitemOpen
  \bibfield  {author} {\bibinfo {author} {\bibfnamefont {H.}~\bibnamefont
  {Terletska}}, \bibinfo {author} {\bibfnamefont {S.-X.}\ \bibnamefont {Yang}},
  \bibinfo {author} {\bibfnamefont {Z.~Y.}\ \bibnamefont {Meng}}, \bibinfo
  {author} {\bibfnamefont {J.}~\bibnamefont {Moreno}},\ and\ \bibinfo {author}
  {\bibfnamefont {M.}~\bibnamefont {Jarrell}},\ }\bibfield  {title} {\bibinfo
  {title} {Dual fermion method for disordered electronic systems},\ }\href
  {https://doi.org/10.1103/PhysRevB.87.134208} {\bibfield  {journal} {\bibinfo
  {journal} {Phys. Rev. B}\ }\textbf {\bibinfo {volume} {87}},\ \bibinfo
  {pages} {134208} (\bibinfo {year} {2013})}\BibitemShut {NoStop}%
\bibitem [{\citenamefont {Yang}\ \emph {et~al.}(2014)\citenamefont {Yang},
  \citenamefont {Haase}, \citenamefont {Terletska}, \citenamefont {Meng},
  \citenamefont {Pruschke}, \citenamefont {Moreno},\ and\ \citenamefont
  {Jarrell}}]{Yang2014}%
  \BibitemOpen
  \bibfield  {author} {\bibinfo {author} {\bibfnamefont {S.-X.}\ \bibnamefont
  {Yang}}, \bibinfo {author} {\bibfnamefont {P.}~\bibnamefont {Haase}},
  \bibinfo {author} {\bibfnamefont {H.}~\bibnamefont {Terletska}}, \bibinfo
  {author} {\bibfnamefont {Z.~Y.}\ \bibnamefont {Meng}}, \bibinfo {author}
  {\bibfnamefont {T.}~\bibnamefont {Pruschke}}, \bibinfo {author}
  {\bibfnamefont {J.}~\bibnamefont {Moreno}},\ and\ \bibinfo {author}
  {\bibfnamefont {M.}~\bibnamefont {Jarrell}},\ }\bibfield  {title} {\bibinfo
  {title} {Dual-fermion approach to interacting disordered fermion systems},\
  }\href {https://doi.org/10.1103/PhysRevB.89.195116} {\bibfield  {journal}
  {\bibinfo  {journal} {Phys. Rev. B}\ }\textbf {\bibinfo {volume} {89}},\
  \bibinfo {pages} {195116} (\bibinfo {year} {2014})}\BibitemShut {NoStop}%
\bibitem [{\citenamefont {Weh}\ \emph {et~al.}(2021)\citenamefont {Weh},
  \citenamefont {Zhang}, \citenamefont {\"Ostlin}, \citenamefont {Terletska},
  \citenamefont {Bauernfeind}, \citenamefont {Tam}, \citenamefont {Evertz},
  \citenamefont {Byczuk}, \citenamefont {Vollhardt},\ and\ \citenamefont
  {Chioncel}}]{Weh2021}%
  \BibitemOpen
  \bibfield  {author} {\bibinfo {author} {\bibfnamefont {A.}~\bibnamefont
  {Weh}}, \bibinfo {author} {\bibfnamefont {Y.}~\bibnamefont {Zhang}}, \bibinfo
  {author} {\bibfnamefont {A.}~\bibnamefont {\"Ostlin}}, \bibinfo {author}
  {\bibfnamefont {H.}~\bibnamefont {Terletska}}, \bibinfo {author}
  {\bibfnamefont {D.}~\bibnamefont {Bauernfeind}}, \bibinfo {author}
  {\bibfnamefont {K.-M.}\ \bibnamefont {Tam}}, \bibinfo {author} {\bibfnamefont
  {H.~G.}\ \bibnamefont {Evertz}}, \bibinfo {author} {\bibfnamefont
  {K.}~\bibnamefont {Byczuk}}, \bibinfo {author} {\bibfnamefont
  {D.}~\bibnamefont {Vollhardt}},\ and\ \bibinfo {author} {\bibfnamefont
  {L.}~\bibnamefont {Chioncel}},\ }\bibfield  {title} {\bibinfo {title}
  {Dynamical mean-field theory of the anderson-hubbard model with local and
  nonlocal disorder in tensor formulation},\ }\href
  {https://doi.org/10.1103/PhysRevB.104.045127} {\bibfield  {journal} {\bibinfo
   {journal} {Phys. Rev. B}\ }\textbf {\bibinfo {volume} {104}},\ \bibinfo
  {pages} {045127} (\bibinfo {year} {2021})}\BibitemShut {NoStop}%
\bibitem [{\citenamefont {Dohner}\ \emph {et~al.}(2022)\citenamefont {Dohner},
  \citenamefont {Terletska}, \citenamefont {Tam}, \citenamefont {Moreno},\ and\
  \citenamefont {Fotso}}]{PhysRevB.106.195156}%
  \BibitemOpen
  \bibfield  {author} {\bibinfo {author} {\bibfnamefont {E.}~\bibnamefont
  {Dohner}}, \bibinfo {author} {\bibfnamefont {H.}~\bibnamefont {Terletska}},
  \bibinfo {author} {\bibfnamefont {K.-M.}\ \bibnamefont {Tam}}, \bibinfo
  {author} {\bibfnamefont {J.}~\bibnamefont {Moreno}},\ and\ \bibinfo {author}
  {\bibfnamefont {H.~F.}\ \bibnamefont {Fotso}},\ }\bibfield  {title} {\bibinfo
  {title} {Nonequilibrium $\text{DMFT}+\text{CPA}$ for correlated disordered
  systems},\ }\href {https://doi.org/10.1103/PhysRevB.106.195156} {\bibfield
  {journal} {\bibinfo  {journal} {Phys. Rev. B}\ }\textbf {\bibinfo {volume}
  {106}},\ \bibinfo {pages} {195156} (\bibinfo {year} {2022})}\BibitemShut
  {NoStop}%
\bibitem [{\citenamefont {Stefanucci}\ and\ \citenamefont
  {Van~Leeuwen}(2013)}]{stefanucci2013nonequilibrium}%
  \BibitemOpen
  \bibfield  {author} {\bibinfo {author} {\bibfnamefont {G.}~\bibnamefont
  {Stefanucci}}\ and\ \bibinfo {author} {\bibfnamefont {R.}~\bibnamefont
  {Van~Leeuwen}},\ }\href@noop {} {\emph {\bibinfo {title} {Nonequilibrium
  many-body theory of quantum systems: a modern introduction}}}\ (\bibinfo
  {publisher} {Cambridge University Press},\ \bibinfo {year}
  {2013})\BibitemShut {NoStop}%
\bibitem [{\citenamefont {Sch{\"u}ler}\ \emph {et~al.}(2020)\citenamefont
  {Sch{\"u}ler}, \citenamefont {Gole{\v z}}, \citenamefont {Murakami},
  \citenamefont {Bittner}, \citenamefont {Herrmann}, \citenamefont {Strand},
  \citenamefont {Werner},\ and\ \citenamefont {Eckstein}}]{SCHULER2020107484}%
  \BibitemOpen
  \bibfield  {author} {\bibinfo {author} {\bibfnamefont {M.}~\bibnamefont
  {Sch{\"u}ler}}, \bibinfo {author} {\bibfnamefont {D.}~\bibnamefont {Gole{\v
  z}}}, \bibinfo {author} {\bibfnamefont {Y.}~\bibnamefont {Murakami}},
  \bibinfo {author} {\bibfnamefont {N.}~\bibnamefont {Bittner}}, \bibinfo
  {author} {\bibfnamefont {A.}~\bibnamefont {Herrmann}}, \bibinfo {author}
  {\bibfnamefont {H.~U.}\ \bibnamefont {Strand}}, \bibinfo {author}
  {\bibfnamefont {P.}~\bibnamefont {Werner}},\ and\ \bibinfo {author}
  {\bibfnamefont {M.}~\bibnamefont {Eckstein}},\ }\bibfield  {title} {\bibinfo
  {title} {Nessi: The non-equilibrium systems simulation package},\ }\href
  {https://doi.org/https://doi.org/10.1016/j.cpc.2020.107484} {\bibfield
  {journal} {\bibinfo  {journal} {Computer Physics Communications}\ }\textbf
  {\bibinfo {volume} {257}},\ \bibinfo {pages} {107484} (\bibinfo {year}
  {2020})}\BibitemShut {NoStop}%
\bibitem [{\citenamefont {Haug}\ \emph {et~al.}(2008)\citenamefont {Haug},
  \citenamefont {Jauho},\ and\ \citenamefont {Cardona}}]{Haug2008}%
  \BibitemOpen
  \bibfield  {author} {\bibinfo {author} {\bibfnamefont {H.}~\bibnamefont
  {Haug}}, \bibinfo {author} {\bibfnamefont {A.-P.}\ \bibnamefont {Jauho}},\
  and\ \bibinfo {author} {\bibfnamefont {M.}~\bibnamefont {Cardona}},\
  }\href@noop {} {\emph {\bibinfo {title} {Quantum kinetics in transport and
  optics of semiconductors}}},\ Vol.~\bibinfo {volume} {2}\ (\bibinfo
  {publisher} {Springer},\ \bibinfo {year} {2008})\BibitemShut {NoStop}%
\bibitem [{\citenamefont {Kamenev}(2011)}]{Kamenev2011}%
  \BibitemOpen
  \bibfield  {author} {\bibinfo {author} {\bibfnamefont {A.}~\bibnamefont
  {Kamenev}},\ }\href@noop {} {\emph {\bibinfo {title} {Field theory of
  non-equilibrium systems}}}\ (\bibinfo  {publisher} {Cambridge University
  Press},\ \bibinfo {year} {2011})\BibitemShut {NoStop}%
\bibitem [{\citenamefont {Turek}\ \emph {et~al.}(2013)\citenamefont {Turek},
  \citenamefont {Drchal}, \citenamefont {Kudrnovsk{\`y}}, \citenamefont {Sob},\
  and\ \citenamefont {Weinberger}}]{turek2013electronic}%
  \BibitemOpen
  \bibfield  {author} {\bibinfo {author} {\bibfnamefont {I.}~\bibnamefont
  {Turek}}, \bibinfo {author} {\bibfnamefont {V.}~\bibnamefont {Drchal}},
  \bibinfo {author} {\bibfnamefont {J.}~\bibnamefont {Kudrnovsk{\`y}}},
  \bibinfo {author} {\bibfnamefont {M.}~\bibnamefont {Sob}},\ and\ \bibinfo
  {author} {\bibfnamefont {P.}~\bibnamefont {Weinberger}},\ }\href@noop {}
  {\emph {\bibinfo {title} {Electronic structure of disordered alloys, surfaces
  and interfaces}}}\ (\bibinfo  {publisher} {Springer Science \& Business
  Media},\ \bibinfo {year} {2013})\BibitemShut {NoStop}%
\bibitem [{\citenamefont {Yan}\ and\ \citenamefont
  {Ke}(2016)}]{PhysRevB.94.045424}%
  \BibitemOpen
  \bibfield  {author} {\bibinfo {author} {\bibfnamefont {J.}~\bibnamefont
  {Yan}}\ and\ \bibinfo {author} {\bibfnamefont {Y.}~\bibnamefont {Ke}},\
  }\bibfield  {title} {\bibinfo {title} {Generalized nonequilibrium vertex
  correction method in coherent medium theory for quantum transport simulation
  of disordered nanoelectronics},\ }\href
  {https://doi.org/10.1103/PhysRevB.94.045424} {\bibfield  {journal} {\bibinfo
  {journal} {Phys. Rev. B}\ }\textbf {\bibinfo {volume} {94}},\ \bibinfo
  {pages} {045424} (\bibinfo {year} {2016})}\BibitemShut {NoStop}%
\bibitem [{\citenamefont {Zhang}\ \emph {et~al.}(1993)\citenamefont {Zhang},
  \citenamefont {Rozenberg},\ and\ \citenamefont {Kotliar}}]{Zhang1993}%
  \BibitemOpen
  \bibfield  {author} {\bibinfo {author} {\bibfnamefont {X.~Y.}\ \bibnamefont
  {Zhang}}, \bibinfo {author} {\bibfnamefont {M.~J.}\ \bibnamefont
  {Rozenberg}},\ and\ \bibinfo {author} {\bibfnamefont {G.}~\bibnamefont
  {Kotliar}},\ }\bibfield  {title} {\bibinfo {title} {Mott transition in the
  d=\ensuremath{\infty} hubbard model at zero temperature},\ }\href
  {https://doi.org/10.1103/PhysRevLett.70.1666} {\bibfield  {journal} {\bibinfo
   {journal} {Phys. Rev. Lett.}\ }\textbf {\bibinfo {volume} {70}},\ \bibinfo
  {pages} {1666} (\bibinfo {year} {1993})}\BibitemShut {NoStop}%
\bibitem [{\citenamefont {Tsuji}\ and\ \citenamefont
  {Werner}(2013)}]{Tsuji2013}%
  \BibitemOpen
  \bibfield  {author} {\bibinfo {author} {\bibfnamefont {N.}~\bibnamefont
  {Tsuji}}\ and\ \bibinfo {author} {\bibfnamefont {P.}~\bibnamefont {Werner}},\
  }\bibfield  {title} {\bibinfo {title} {Nonequilibrium dynamical mean-field
  theory based on weak-coupling perturbation expansions: Application to
  dynamical symmetry breaking in the hubbard model},\ }\href
  {https://doi.org/10.1103/PhysRevB.88.165115} {\bibfield  {journal} {\bibinfo
  {journal} {Phys. Rev. B}\ }\textbf {\bibinfo {volume} {88}},\ \bibinfo
  {pages} {165115} (\bibinfo {year} {2013})}\BibitemShut {NoStop}%
\bibitem [{\citenamefont {Baym}(1962)}]{PhysRev.127.1391}%
  \BibitemOpen
  \bibfield  {author} {\bibinfo {author} {\bibfnamefont {G.}~\bibnamefont
  {Baym}},\ }\bibfield  {title} {\bibinfo {title} {Self-consistent
  approximations in many-body systems},\ }\href
  {https://doi.org/10.1103/PhysRev.127.1391} {\bibfield  {journal} {\bibinfo
  {journal} {Phys. Rev.}\ }\textbf {\bibinfo {volume} {127}},\ \bibinfo {pages}
  {1391} (\bibinfo {year} {1962})}\BibitemShut {NoStop}%
\bibitem [{\citenamefont {Baym}\ and\ \citenamefont
  {Kadanoff}(1961)}]{PhysRev.124.287}%
  \BibitemOpen
  \bibfield  {author} {\bibinfo {author} {\bibfnamefont {G.}~\bibnamefont
  {Baym}}\ and\ \bibinfo {author} {\bibfnamefont {L.~P.}\ \bibnamefont
  {Kadanoff}},\ }\bibfield  {title} {\bibinfo {title} {Conservation laws and
  correlation functions},\ }\href {https://doi.org/10.1103/PhysRev.124.287}
  {\bibfield  {journal} {\bibinfo  {journal} {Phys. Rev.}\ }\textbf {\bibinfo
  {volume} {124}},\ \bibinfo {pages} {287} (\bibinfo {year}
  {1961})}\BibitemShut {NoStop}%
\bibitem [{\citenamefont {Eckstein}\ \emph {et~al.}(2010)\citenamefont
  {Eckstein}, \citenamefont {Kollar},\ and\ \citenamefont
  {Werner}}]{Eckstein2010}%
  \BibitemOpen
  \bibfield  {author} {\bibinfo {author} {\bibfnamefont {M.}~\bibnamefont
  {Eckstein}}, \bibinfo {author} {\bibfnamefont {M.}~\bibnamefont {Kollar}},\
  and\ \bibinfo {author} {\bibfnamefont {P.}~\bibnamefont {Werner}},\
  }\bibfield  {title} {\bibinfo {title} {Interaction quench in the hubbard
  model: Relaxation of the spectral function and the optical conductivity},\
  }\href {https://doi.org/10.1103/PhysRevB.81.115131} {\bibfield  {journal}
  {\bibinfo  {journal} {Phys. Rev. B}\ }\textbf {\bibinfo {volume} {81}},\
  \bibinfo {pages} {115131} (\bibinfo {year} {2010})}\BibitemShut {NoStop}%
\bibitem [{\citenamefont {Jauho}\ \emph {et~al.}(1994)\citenamefont {Jauho},
  \citenamefont {Wingreen},\ and\ \citenamefont {Meir}}]{Jauho1994}%
  \BibitemOpen
  \bibfield  {author} {\bibinfo {author} {\bibfnamefont {A.-P.}\ \bibnamefont
  {Jauho}}, \bibinfo {author} {\bibfnamefont {N.~S.}\ \bibnamefont
  {Wingreen}},\ and\ \bibinfo {author} {\bibfnamefont {Y.}~\bibnamefont
  {Meir}},\ }\bibfield  {title} {\bibinfo {title} {Time-dependent transport in
  interacting and noninteracting resonant-tunneling systems},\ }\href
  {https://doi.org/10.1103/PhysRevB.50.5528} {\bibfield  {journal} {\bibinfo
  {journal} {Phys. Rev. B}\ }\textbf {\bibinfo {volume} {50}},\ \bibinfo
  {pages} {5528} (\bibinfo {year} {1994})}\BibitemShut {NoStop}%
\bibitem [{\citenamefont {Aron}\ \emph {et~al.}(2012)\citenamefont {Aron},
  \citenamefont {Kotliar},\ and\ \citenamefont {Weber}}]{Aron2012}%
  \BibitemOpen
  \bibfield  {author} {\bibinfo {author} {\bibfnamefont {C.}~\bibnamefont
  {Aron}}, \bibinfo {author} {\bibfnamefont {G.}~\bibnamefont {Kotliar}},\ and\
  \bibinfo {author} {\bibfnamefont {C.}~\bibnamefont {Weber}},\ }\bibfield
  {title} {\bibinfo {title} {Dimensional crossover driven by an electric
  field},\ }\href {https://doi.org/10.1103/PhysRevLett.108.086401} {\bibfield
  {journal} {\bibinfo  {journal} {Phys. Rev. Lett.}\ }\textbf {\bibinfo
  {volume} {108}},\ \bibinfo {pages} {086401} (\bibinfo {year}
  {2012})}\BibitemShut {NoStop}%
\bibitem [{\citenamefont {Li}\ \emph {et~al.}(2015)\citenamefont {Li},
  \citenamefont {Aron}, \citenamefont {Kotliar},\ and\ \citenamefont
  {Han}}]{Li2015}%
  \BibitemOpen
  \bibfield  {author} {\bibinfo {author} {\bibfnamefont {J.}~\bibnamefont
  {Li}}, \bibinfo {author} {\bibfnamefont {C.}~\bibnamefont {Aron}}, \bibinfo
  {author} {\bibfnamefont {G.}~\bibnamefont {Kotliar}},\ and\ \bibinfo {author}
  {\bibfnamefont {J.~E.}\ \bibnamefont {Han}},\ }\bibfield  {title} {\bibinfo
  {title} {Electric-field-driven resistive switching in the dissipative hubbard
  model},\ }\href {https://doi.org/10.1103/PhysRevLett.114.226403} {\bibfield
  {journal} {\bibinfo  {journal} {Phys. Rev. Lett.}\ }\textbf {\bibinfo
  {volume} {114}},\ \bibinfo {pages} {226403} (\bibinfo {year}
  {2015})}\BibitemShut {NoStop}%
\bibitem [{\citenamefont {Li}\ and\ \citenamefont {Eckstein}(2021)}]{Li2021}%
  \BibitemOpen
  \bibfield  {author} {\bibinfo {author} {\bibfnamefont {J.}~\bibnamefont
  {Li}}\ and\ \bibinfo {author} {\bibfnamefont {M.}~\bibnamefont {Eckstein}},\
  }\bibfield  {title} {\bibinfo {title} {Nonequilibrium steady-state theory of
  photodoped mott insulators},\ }\href
  {https://doi.org/10.1103/PhysRevB.103.045133} {\bibfield  {journal} {\bibinfo
   {journal} {Phys. Rev. B}\ }\textbf {\bibinfo {volume} {103}},\ \bibinfo
  {pages} {045133} (\bibinfo {year} {2021})}\BibitemShut {NoStop}%
\bibitem [{\citenamefont {Yan}\ and\ \citenamefont {Jani\ifmmode~\check{s}\else
  \v{s}\fi{}}(2022)}]{PhysRevB.105.085122}%
  \BibitemOpen
  \bibfield  {author} {\bibinfo {author} {\bibfnamefont {J.}~\bibnamefont
  {Yan}}\ and\ \bibinfo {author} {\bibfnamefont {V.}~\bibnamefont
  {Jani\ifmmode~\check{s}\else \v{s}\fi{}}},\ }\bibfield  {title} {\bibinfo
  {title} {Single-impurity anderson model out of equilibrium: A two-particle
  semianalytic approach},\ }\href {https://doi.org/10.1103/PhysRevB.105.085122}
  {\bibfield  {journal} {\bibinfo  {journal} {Phys. Rev. B}\ }\textbf {\bibinfo
  {volume} {105}},\ \bibinfo {pages} {085122} (\bibinfo {year}
  {2022})}\BibitemShut {NoStop}%
\bibitem [{Note1()}]{Note1}%
  \BibitemOpen
  \bibinfo {note} {For continuous distributed disorders, one can simulate this
  by sampling over the distribution function and transforming to the
  multi-component alloy problem.}\BibitemShut {Stop}%
\bibitem [{Note2()}]{Note2}%
  \BibitemOpen
  \bibinfo {note} {The parameters are chosen to be appropriate for
  half-filling, since IPT solver gives reasonable results in this
  regime.}\BibitemShut {Stop}%
\bibitem [{\citenamefont {Zhou}\ \emph {et~al.}(2016)\citenamefont {Zhou},
  \citenamefont {Chen},\ and\ \citenamefont {Guo}}]{PhysRevB.94.075426}%
  \BibitemOpen
  \bibfield  {author} {\bibinfo {author} {\bibfnamefont {C.}~\bibnamefont
  {Zhou}}, \bibinfo {author} {\bibfnamefont {X.}~\bibnamefont {Chen}},\ and\
  \bibinfo {author} {\bibfnamefont {H.}~\bibnamefont {Guo}},\ }\bibfield
  {title} {\bibinfo {title} {Theory of quantum transport in disordered systems
  driven by voltage pulse},\ }\href
  {https://doi.org/10.1103/PhysRevB.94.075426} {\bibfield  {journal} {\bibinfo
  {journal} {Phys. Rev. B}\ }\textbf {\bibinfo {volume} {94}},\ \bibinfo
  {pages} {075426} (\bibinfo {year} {2016})}\BibitemShut {NoStop}%
\bibitem [{Note3()}]{Note3}%
  \BibitemOpen
  \bibinfo {note} {Our calculation uses an inverse temperature of $\beta =
  0.05$ instead of zero temperature as in Ref.~\protect \rev@citealp
  {PhysRevB.94.075426}, but this does not significantly affect the
  results.}\BibitemShut {Stop}%
\bibitem [{\citenamefont {Werner}\ \emph {et~al.}(2010)\citenamefont {Werner},
  \citenamefont {Oka}, \citenamefont {Eckstein},\ and\ \citenamefont
  {Millis}}]{Werner2010}%
  \BibitemOpen
  \bibfield  {author} {\bibinfo {author} {\bibfnamefont {P.}~\bibnamefont
  {Werner}}, \bibinfo {author} {\bibfnamefont {T.}~\bibnamefont {Oka}},
  \bibinfo {author} {\bibfnamefont {M.}~\bibnamefont {Eckstein}},\ and\
  \bibinfo {author} {\bibfnamefont {A.~J.}\ \bibnamefont {Millis}},\ }\bibfield
   {title} {\bibinfo {title} {Weak-coupling quantum monte carlo calculations on
  the keldysh contour: Theory and application to the current-voltage
  characteristics of the anderson model},\ }\href
  {https://doi.org/10.1103/PhysRevB.81.035108} {\bibfield  {journal} {\bibinfo
  {journal} {Phys. Rev. B}\ }\textbf {\bibinfo {volume} {81}},\ \bibinfo
  {pages} {035108} (\bibinfo {year} {2010})}\BibitemShut {NoStop}%
\bibitem [{\citenamefont {Eckel}\ \emph {et~al.}(2010)\citenamefont {Eckel},
  \citenamefont {Heidrich-Meisner}, \citenamefont {Jakobs}, \citenamefont
  {Thorwart}, \citenamefont {Pletyukhov},\ and\ \citenamefont
  {Egger}}]{Eckel2010}%
  \BibitemOpen
  \bibfield  {author} {\bibinfo {author} {\bibfnamefont {J.}~\bibnamefont
  {Eckel}}, \bibinfo {author} {\bibfnamefont {F.}~\bibnamefont
  {Heidrich-Meisner}}, \bibinfo {author} {\bibfnamefont {S.~G.}\ \bibnamefont
  {Jakobs}}, \bibinfo {author} {\bibfnamefont {M.}~\bibnamefont {Thorwart}},
  \bibinfo {author} {\bibfnamefont {M.}~\bibnamefont {Pletyukhov}},\ and\
  \bibinfo {author} {\bibfnamefont {R.}~\bibnamefont {Egger}},\ }\bibfield
  {title} {\bibinfo {title} {Comparative study of theoretical methods for
  non-equilibrium quantum transport},\ }\href
  {https://doi.org/10.1088/1367-2630/12/4/043042} {\bibfield  {journal}
  {\bibinfo  {journal} {New Journal of Physics}\ }\textbf {\bibinfo {volume}
  {12}},\ \bibinfo {pages} {043042} (\bibinfo {year} {2010})}\BibitemShut
  {NoStop}%
\bibitem [{\citenamefont {Meir}\ and\ \citenamefont
  {Wingreen}(1992)}]{Meir1992}%
  \BibitemOpen
  \bibfield  {author} {\bibinfo {author} {\bibfnamefont {Y.}~\bibnamefont
  {Meir}}\ and\ \bibinfo {author} {\bibfnamefont {N.~S.}\ \bibnamefont
  {Wingreen}},\ }\bibfield  {title} {\bibinfo {title} {Landauer formula for the
  current through an interacting electron region},\ }\href
  {https://doi.org/10.1103/PhysRevLett.68.2512} {\bibfield  {journal} {\bibinfo
   {journal} {Phys. Rev. Lett.}\ }\textbf {\bibinfo {volume} {68}},\ \bibinfo
  {pages} {2512} (\bibinfo {year} {1992})}\BibitemShut {NoStop}%
\bibitem [{\citenamefont {Rubtsov}\ \emph {et~al.}(2005)\citenamefont
  {Rubtsov}, \citenamefont {Savkin},\ and\ \citenamefont
  {Lichtenstein}}]{Rubtsov2005}%
  \BibitemOpen
  \bibfield  {author} {\bibinfo {author} {\bibfnamefont {A.~N.}\ \bibnamefont
  {Rubtsov}}, \bibinfo {author} {\bibfnamefont {V.~V.}\ \bibnamefont
  {Savkin}},\ and\ \bibinfo {author} {\bibfnamefont {A.~I.}\ \bibnamefont
  {Lichtenstein}},\ }\bibfield  {title} {\bibinfo {title} {Continuous-time
  quantum monte carlo method for fermions},\ }\href
  {https://doi.org/10.1103/PhysRevB.72.035122} {\bibfield  {journal} {\bibinfo
  {journal} {Phys. Rev. B}\ }\textbf {\bibinfo {volume} {72}},\ \bibinfo
  {pages} {035122} (\bibinfo {year} {2005})}\BibitemShut {NoStop}%
\bibitem [{\citenamefont {Werner}\ \emph {et~al.}(2006)\citenamefont {Werner},
  \citenamefont {Comanac}, \citenamefont {de' Medici}, \citenamefont {Troyer},\
  and\ \citenamefont {Millis}}]{Werner2006}%
  \BibitemOpen
  \bibfield  {author} {\bibinfo {author} {\bibfnamefont {P.}~\bibnamefont
  {Werner}}, \bibinfo {author} {\bibfnamefont {A.}~\bibnamefont {Comanac}},
  \bibinfo {author} {\bibfnamefont {L.}~\bibnamefont {de' Medici}}, \bibinfo
  {author} {\bibfnamefont {M.}~\bibnamefont {Troyer}},\ and\ \bibinfo {author}
  {\bibfnamefont {A.~J.}\ \bibnamefont {Millis}},\ }\bibfield  {title}
  {\bibinfo {title} {Continuous-time solver for quantum impurity models},\
  }\href {https://doi.org/10.1103/PhysRevLett.97.076405} {\bibfield  {journal}
  {\bibinfo  {journal} {Phys. Rev. Lett.}\ }\textbf {\bibinfo {volume} {97}},\
  \bibinfo {pages} {076405} (\bibinfo {year} {2006})}\BibitemShut {NoStop}%
\bibitem [{\citenamefont {Eckstein}\ and\ \citenamefont
  {Werner}(2010)}]{Eckstein2010b}%
  \BibitemOpen
  \bibfield  {author} {\bibinfo {author} {\bibfnamefont {M.}~\bibnamefont
  {Eckstein}}\ and\ \bibinfo {author} {\bibfnamefont {P.}~\bibnamefont
  {Werner}},\ }\bibfield  {title} {\bibinfo {title} {Nonequilibrium dynamical
  mean-field calculations based on the noncrossing approximation and its
  generalizations},\ }\href {https://doi.org/10.1103/PhysRevB.82.115115}
  {\bibfield  {journal} {\bibinfo  {journal} {Phys. Rev. B}\ }\textbf {\bibinfo
  {volume} {82}},\ \bibinfo {pages} {115115} (\bibinfo {year}
  {2010})}\BibitemShut {NoStop}%
\bibitem [{\citenamefont {Maciejko}\ \emph {et~al.}(2006)\citenamefont
  {Maciejko}, \citenamefont {Wang},\ and\ \citenamefont
  {Guo}}]{PhysRevB.74.085324}%
  \BibitemOpen
  \bibfield  {author} {\bibinfo {author} {\bibfnamefont {J.}~\bibnamefont
  {Maciejko}}, \bibinfo {author} {\bibfnamefont {J.}~\bibnamefont {Wang}},\
  and\ \bibinfo {author} {\bibfnamefont {H.}~\bibnamefont {Guo}},\ }\bibfield
  {title} {\bibinfo {title} {Time-dependent quantum transport far from
  equilibrium: An exact nonlinear response theory},\ }\href
  {https://doi.org/10.1103/PhysRevB.74.085324} {\bibfield  {journal} {\bibinfo
  {journal} {Phys. Rev. B}\ }\textbf {\bibinfo {volume} {74}},\ \bibinfo
  {pages} {085324} (\bibinfo {year} {2006})}\BibitemShut {NoStop}%
\bibitem [{\citenamefont {Godfrin}(1991)}]{Godfrin1991}%
  \BibitemOpen
  \bibfield  {author} {\bibinfo {author} {\bibfnamefont {E.~M.}\ \bibnamefont
  {Godfrin}},\ }\bibfield  {title} {\bibinfo {title} {A method to compute the
  inverse of an n-block tridiagonal quasi-hermitian matrix},\ }\href
  {http://stacks.iop.org/0953-8984/3/i=40/a=005} {\bibfield  {journal}
  {\bibinfo  {journal} {Journal of Physics: Condensed Matter}\ }\textbf
  {\bibinfo {volume} {3}},\ \bibinfo {pages} {7843} (\bibinfo {year}
  {1991})}\BibitemShut {NoStop}%
\bibitem [{\citenamefont {Sancho}\ \emph {et~al.}(1985)\citenamefont {Sancho},
  \citenamefont {Sancho}, \citenamefont {Sancho},\ and\ \citenamefont
  {Rubio}}]{Sancho1985}%
  \BibitemOpen
  \bibfield  {author} {\bibinfo {author} {\bibfnamefont {M.~P.~L.}\
  \bibnamefont {Sancho}}, \bibinfo {author} {\bibfnamefont {J.~M.~L.}\
  \bibnamefont {Sancho}}, \bibinfo {author} {\bibfnamefont {J.~M.~L.}\
  \bibnamefont {Sancho}},\ and\ \bibinfo {author} {\bibfnamefont
  {J.}~\bibnamefont {Rubio}},\ }\bibfield  {title} {\bibinfo {title} {Highly
  convergent schemes for the calculation of bulk and surface green functions},\
  }\href {https://doi.org/10.1088/0305-4608/15/4/009} {\bibfield  {journal}
  {\bibinfo  {journal} {Journal of Physics F: Metal Physics}\ }\textbf
  {\bibinfo {volume} {15}},\ \bibinfo {pages} {851} (\bibinfo {year}
  {1985})}\BibitemShut {NoStop}%
\end{thebibliography}%

\end{document}